\documentclass[10pt,journal,compsoc]{IEEEtran}
\usepackage{graphicx}
\usepackage{subfigure}
\usepackage{amsfonts}
\usepackage{textcomp}
\usepackage{caption}
\usepackage{url}
\usepackage{amsmath}
\usepackage{mathtools}
\usepackage{color}
\usepackage{flushend}
\usepackage{caption,algpseudocode,algorithm,setspace}
\usepackage{scalerel,stackengine}
\stackMath
\newcommand\reallywidehat[1]{%
\savestack{\tmpbox}{\stretchto{%
  \scaleto{%
    \scalerel*[\widthof{\ensuremath{#1}}]{\kern-.6pt\bigwedge\kern-.6pt}%
    {\rule[-\textheight/2]{1ex}{\textheight}}
  }{\textheight}%
}{0.5ex}}%
\stackon[1pt]{#1}{\tmpbox}%
}
\usepackage{booktabs} 
\usepackage{listings}

\ifCLASSINFOpdf
\else
\fi
\begin{document}
%
\title{ExaGeoStat: A High Performance Unified Software for Geostatistics on Manycore Systems}
\author{Sameh~Abdulah,
	Hatem~Ltaief,
	Ying~Sun,
	Marc~G. Genton,
	and~David~E. Keyes
	\IEEEcompsocitemizethanks{\IEEEcompsocthanksitem King Abdullah University of Science and Technology (KAUST), Extreme Computing Research Center, Computer, Electrical, and Mathematical Sciences and Engineering Division (CEMSE), Thuwal, 23955-6900, Saudi Arabia.\protect\\
	E-mail: Sameh.Abdulah@kaust.edu.sa}
\thanks{Manuscript received M DD, YYYY; revised M DD, YYYY.}}

%
%

\markboth{IEEE TRANSACTIONS ON PARALLEL AND DISTRIBUTED SYSTEMS,~Vol.~XX, No.~X, M~YYYY}%
{Shell \MakeLowercase{\textit{et al.}}: ExaGeoStat: A High-Performance Unified Software for Geostatistics on Manycore Systems}
%



\IEEEtitleabstractindextext{%

	\begin{abstract}
		We present \emph{ExaGeoStat}, a high performance software for geospatial statistics in 
		climate and environment modeling. In contrast to simulation based on partial differential equations
		derived from first-principles modeling, \emph{ExaGeoStat} employs a 
		statistical model based on the evaluation of the Gaussian log-likelihood function, 
		which operates on a large dense covariance matrix. Generated by the parametrizable
		Mat\'{e}rn covariance function, 
		the resulting matrix is symmetric and positive definite. 
		The computational tasks involved during the evaluation of the 
		Gaussian log-likelihood function become daunting as the number $n$ of geographical locations
		grows, as ${\mathcal O}(n^2)$ storage and ${\mathcal O}(n^3)$ 
		operations are required.  
		While many approximation methods have been devised from the side of statistical modeling to 
		ameliorate these polynomial complexities, we are interested here in the complementary
		approach of evaluating the exact algebraic result by exploiting advances in solution algorithms
		and many-core computer architectures.
		Using state-of-the-art high performance 
		dense linear algebra libraries associated with various leading edge 
		parallel architectures (Intel KNLs, NVIDIA GPUs, and distributed-memory systems), 
		\emph{ExaGeoStat} raises the game for statistical  applications 
		from climate and environmental science. \emph{ExaGeoStat} provides a 
		reference evaluation of statistical parameters, with which to assess the
		validity of the various approaches based on approximation.
		The software takes a first step in the merger of large-scale data analytics and 
		extreme computing for geospatial statistical applications, to be followed by 
		additional complexity reducing improvements from the solver side that can be
		implemented under the same interface. Thus, a single uncompromised statistical model
		can ultimately be executed in a wide variety of emerging exascale environments.
	\end{abstract}
	\begin{IEEEkeywords}
		MAXIMUM LIKELIHOOD OPTIMIZATION; MAT\'{E}RN COVARIANCE FUNCTION; HIGH PERFORMANCE COMPUTING; CLIMATE/ENVIRONMENT APPLICATIONS; PREDICTION.
\end{IEEEkeywords}}


\maketitle

\section{Introduction}
\label{sec:intro}
Big data applications and traditional high performance-oriented computing have followed
independent paths to the present, but important opportunities now arise that
can be addressed by merging the two.  As a prominent big data application, geospatial
statistics is increasingly performance-bound.
This paper describes the Exascale GeoStatistics (\emph{ExaGeoStat}) software, a high-performance, unified 
software for geostatistics on manycore systems, which targets 
climate and environment prediction applications using techniques from geospatial
statistics. We believe that such a software may play an important role  at the 
intersection of big data and extreme computing by allowing applications with prohibitively large memory 
footprints to be deployed at the desired
scale on modern hardware architectures, exploiting recent
software developments in computational linear algebra. \emph{ExaGeoStat} is intended 
to bridge the aforementioned gap, attracting the geospatial statistics community to the vast potential of 
high-performance computing and providing fresh inspiration for algorithm and software developments
to the HPC community. 

Applications for climate and environmental predictions are among the principal 
simulation workloads running on today's supercomputer facilities.These applications
usually approximate state variables by relying on numerical models to solve a complex
set of partial differential equations, which are based on a combination of first-principles
and empirical models tuned by known measurements, on a highly resolved spatial and temporal grid.
Then, the large volume of results this method produces is post-processed to estimate the quantities of interest.
Such an approach translates the original
big data problem into an HPC-oriented problem, by relying on 
PDE solvers to extract performance on the targeted architectures. 
Instead, \emph{ExaGeoStat} employs a compute-intensive statistical 
model based on the evaluation of the Gaussian 
log-likelihood function, which operates on a large dense covariance matrix. The matrix is generated
directly from the application datasets, using the 
parametrizable Mat\'{e}rn covariance function. The resulting covariance matrix 
is symmetric and positive-definite. 
The computational tasks involved during the evaluation of the 
Gaussian log-likelihood function become daunting as the number $n$ of geographical locations
grows, as ${\mathcal O}(n^2)$ storage and ${\mathcal O}(n^3)$ 
operations are required.

\emph{ExaGeoStat}'s primary goal is not to resolve this complexity challenge \emph{per se}, but to delay
its scaling limitation impact, by maximizing the computational power of emerging architectures. 
The unified software permits to explore 
the computational limits using state-of-the-art high-performance 
dense linear algebra libraries by leveraging a single source code to 
run on various cutting-edge 
parallel architectures, e.g., Intel Xeon, Intel manycore Xeon Phi 
Knights Landing chip (KNL), NVIDIA GPU
accelerators, and distributed-memory homogeneous systems. To achieve this 
software productivity, we rely on the 
dense linear algebra library Chameleon~\cite{chameleon-soft}, which breaks down the tasks of the
traditional bulk-synchronous programming model of LAPACK~\cite{anderson1999lapack} and renders them
for an asynchronous task-based programming model. Task-based programming models have received significant 
attention in computational science and engineering, since they may achieve greater concurrency and mitigate
communication overhead, thus presenting a path to the exascale era~\cite{AgulloBGL15,berry-jcp,dongarra2011exacall}.
Once a numerical algorithm has been expressed in tasks linked by input-output data dependencies,
We use the StarPU dynamic runtime system~\cite{AugThiNamWac11CCPE}  to schedule
the various tasks on the underlying hardware resources. The simulation code need only be written once
since StarPU allows porting to its supported architectures.  
\emph{ExaGeoStat} may thus positively impact the day-to-day simulation work 
of end users 
by efficiently implementing the limiting linear algebra operations on large datasets.

To highlight the software contributions and to verify that the model can be applied to
geostatistical applications, we design a synthetic dataset generator, 
which allows us not only to test the software infrastructure, but 
also to stress the statistical model accordingly. In addition, we experiment
using a soil moisture dataset from the Mississippi River basin.
Although we focus only on soil moisture, 
our software is able to analyze other variables that commonly employ 
the Gaussian log-likelihood function and its flexible Mat\'{e}rn covariance, 
such as temperature, wind speed, etc.  The distillate of this work is two packages that are publicly released as an open-source under
BSD 3-Clause license: \emph{ExaGeoStat} C library\footnote{ExaGeoStat is  available at https://github.com/ecrc/exageostat} 
and  R-wrapper library\footnote{ExaGeoStatR is  available at https://github.com/ecrc/exageostatr}.

The remainder of the paper is organized as follows.
Section~\ref{sec:pb} states the problem, describes related work,
describes the construction of the climate and environment modeling simulation,
and shows how to predict missing measurements  using this constructed model in which we apply a geostatistical approach to compute large dense covariance matrix.
Section~\ref{sec:contrib} highlights our contributions.
Section~\ref{sec:geospat} presents a case study
from a large geographic region, the Mississippi River basin, and notes the effects of 
some alternative representations
of distance in this context.
Section~\ref{sec:sotadla} reviews the dense
linear algebra libraries.
Section~\ref{sec:impl} outlines the geostatistical algorithm, 
as implemented in the \emph{ExaGeoStat} software,
and lays out the overall software stack. Performance
results and analysis are presented in Section~\ref{sec:perf}, using the synthetic 
and the real datasets, and we conclude in Section~\ref{sec:summary}.
\section{Problem Statement}
\label{sec:pb}
Applications in climate and environmental science often deal with a very large number
of measurements regularly or irregularly located across a geographical region. 
In geostatistics, these data are usually modeled as a realization from a  Gaussian spatial random field. 
Specifically, let ${\bf s}_1,\ldots,{\bf s}_n$ denote $n$ spatial locations in $\Bbb{R}^d$, $d\geq 1$,
and let ${\bf Z}=\{Z({\bf s}_1),\ldots,Z({\bf s}_n)\}^\top$ be a realization 
of a Gaussian random field $Z({\bf s})$ 
at those $n$ locations. For simplicity, assume the random field $Z({\bf s})$ 
has a mean zero and stationary parametric covariance function 
$C({\bf h};{\boldsymbol \theta})=\mbox{cov}\{Z({\bf s}),Z({\bf s}+{\bf h})\}$,
where ${\bf h}\in\Bbb{R}^d$ is a spatial lag vector and ${\boldsymbol \theta}\in\Bbb{R}^q$ is
an unknown parameter vector of interest. 
Denote by ${\boldsymbol \Sigma}({\boldsymbol \theta})$ the covariance matrix with
entries ${\boldsymbol \Sigma_{ij}}=C({\bf s}_i-{\bf s}_j;{\boldsymbol \theta})$, $i,j=1,\ldots,n$.
The matrix ${\boldsymbol \Sigma}({\boldsymbol \theta})$ is symmetric and positive definite.
Statistical inference about $\boldsymbol \theta$ is often based on the Gaussian
log-likelihood function:
\begin{equation}
	\label{eq:likeli}
	\ell({\boldsymbol \theta})=-\frac{n}{2}\log(2\pi) - \frac{1}{2}\log |{{\boldsymbol \Sigma}({\boldsymbol \theta})}|-\frac{1}{2}{\bf Z}^\top {\boldsymbol \Sigma}({\boldsymbol \theta})^{-1}{\bf Z}.
\end{equation}
The maximum likelihood estimator of ${\boldsymbol \theta}$ is the value 
$\widehat{\boldsymbol \theta}$ that maximizes (\ref{eq:likeli}). 
When the sample size of $n$ locations
is large and the locations are irregularly spaced,
the evaluation of (\ref{eq:likeli}) becomes challenging because the 
linear solver and log-determinant involving the $n$-by-$n$ dense and unstructured covariance matrix
${\boldsymbol \Sigma}({\boldsymbol \theta})$ requires ${\mathcal O}(n^3)$ floating-point operations
on ${\mathcal O}(n^2)$ memory. For example, assuming a dataset approximately on a grid with $10^3$
 longitude values and $10^3$ latitude values, 
the total number of locations will be $10^6$. Using double-precision floating-point arithmetic, 
the total required memory footprint is 
$10^{12} \times 8 $ bytes $ \sim 8 $ TB. 
The corresponding complexity order is $10^{18}$.
\subsection{Related Work}
In recent years, a large amount of research has been devoted to addressing the 
aforementioned challenge through various approximations; for example, 
covariance tapering~\cite{Furrer2006, sang2012full}, likelihood 
approximations in both the spatial~\cite{Stein:Chi:Wetly:2004} and spectral~\cite{Fuentes:2007} domains, 
latent processes such as Gaussian predictive processes and fixed rank kriging
~\cite{Banerjee:Gelfand:Finley:Sang:2008, Cressie:Johannesson:2008}, and Gaussian Markov random field 
approximations~\cite{fuglstad2015does,lindgren2011explicit, rue2005gaussian, rue2002fitting}; see Sun~\cite{Sun2012} for a review. 
Stein~\cite{stein2013statistical} showed that covariance tapering sometimes performs even worse than assuming 
independent blocks in the covariance; Stein~\cite{stein2014limitations} discussed 
the limitations of low rank approximations; and Markov models  depend on 
the measurements locations, which must be aligned on a fine grid with estimations of the missing values~\cite{sun2016statistically}. 
Very recent methods include the nearest-neighbor Gaussian process models~\cite{datta2016hierarchical},
multiresolution Gaussian process 
models~\cite{nychka2015multiresolution}, equivalent kriging~\cite{kleiber2015equivalent}, 
multi-level restricted Gaussian maximum likelihood estimators~\cite{Castrillon16}, 
and hierarchical low rank representations~\cite{huang2017hierarchical}. However, all these methods 
reduce the computational cost by either approximating the maximum likelihood estimator, 
or by using approximate models that may or may not allow for exact computations. In this paper, 
we propose exploring the computational limits of the {\it exact} evaluation of the 
Gaussian log-likelihood function, i.e., Equation~(\ref{eq:likeli}) with high-performance computing 
and implementing modern techniques to solve these fundamental computational problems in geostatistics.

\subsection{Mat\'{e}rn Covariance Functions}
To construct the covariance matrix ${\boldsymbol \Sigma}({\boldsymbol \theta})$ in Equation
(\ref{eq:likeli}), a valid (positive definite) parametric covariance model is needed. 
Among the many possible covariance models in the literature, the Mat\'{e}rn family~\cite{Matern1986a} has gained widespread interest
in recent years due to its flexibility. The class of Mat\'{e}rn covariance functions~\cite{Handcock1993a}
is widely used in geostatistics and spatial statistics~\cite{chiles2009geostatistics},
machine learning~\cite{BoermGarcke2007}, 
image analysis, weather forecasting and climate science. 
Handcock and Stein~\cite{Handcock1993a} introduced the Mat\'{e}rn form of spatial correlations 
into statistics as a flexible parametric class where one parameter determines
the smoothness of the underlying spatial random field. The history of this 
family of models can be found in~\cite{Guttorp2006a}.
The Mat\'{e}rn form also naturally describes the correlation among temperature 
fields that can be explained by simple energy balance climate models~\cite{North2011a}. 
The Mat\'{e}rn class of covariance functions is defined as
\begin{equation}
	\label{eq:MaternCov}
	C(r;{\boldsymbol \theta})=\frac{\theta_1}{2^{\theta_3-1}\Gamma(\theta_3)}\left(\frac{r}{\theta_2}\right)^{\theta_3} {\mathcal K}_{\theta_3}\left(\frac{r}{\theta_2}\right),
\end{equation}
where $r=\|{\bf s}-{\bf s}'\|$ is the distance between two spatial locations, ${\bf s}$ and ${\bf s}'$, and 
${\boldsymbol \theta}=(\theta_1,\theta_2,\theta_3)^\top$.  Here $\theta_1>0$ is the variance, 
$\theta_2>0$ is a spatial range parameter that measures how quickly the correlation of the random 
field decays with distance, and $\theta_3>0$ controls the smoothness of the 
random field, with larger values of $\theta_3$ corresponding to smoother fields. 

The function ${\mathcal K}_{\theta_3}$ denotes the modified Bessel function of 
the second kind of order $\theta_3$. When $\theta_3=1/2$, the Mat\'{e}rn 
covariance function reduces to the exponential covariance model 
$C(r;{\boldsymbol \theta})=\theta_1 \exp(-r/\theta_2)$, and describes a 
rough field, whereas when $\theta_3=1$, the Mat\'{e}rn covariance function reduces to the Whittle covariance model 
$C(r;{\boldsymbol \theta})=\theta_1 (r/\theta_2){\mathcal K}_1(r/\theta_2)$, 
and describes a smooth field.
The value $\theta_3=\infty$ corresponds to a Gaussian covariance model,
which describes a very smooth field infinitely mean-square differentiable. 
Realizations from a random field with Mat\'{e}rn covariance functions are 
$\lfloor \theta_3-1 \rfloor$ times mean-square differentiable. Thus, the parameter $\theta_3$ 
is used to control the degree of smoothness of the random field. 

In theory, the three parameters of the Mat\'{e}rn covariance function need to be positive real
numbers, but empirical values derived from the empirical covariance of the data can serve
as starting values and provide bounds for the optimization. Moreover, the parameter
$\theta_3$ is rarely found to be larger than 1 or 2 in geophysical applications, as those already correspond to very smooth realizations.

\subsection{Prediction}
The quality of statistical forecasts could be improved by accurately estimating the unknown
parameters of a statistical model. With the aid of a given geospatial data
and measurements, the constructed statistical model is able to
predict missing measurements at new spatial locations.

Assuming unknown measurements vector $\mathbf{Z}_1$ with size $m$  and  know measurements vector $\mathbf{Z}_2$ with size $n$,
the prediction problem can be represented as a multivariate normal
joint distribution as follows~\cite{cressie2015statistics, genton2007separable}

\begin{equation}
	\begin{bmatrix}
		\mathbf{Z}_1 \\
		\mathbf{Z}_2 \\
	\end{bmatrix}
	\sim
	N_{m+n}
	(
	\begin{bmatrix}
		\boldsymbol{\mu}_{1} \\
		\boldsymbol{\mu}_{2} \\
	\end{bmatrix}
	,
	\begin{bmatrix}
		\mathbf{\Sigma}_{11} & \mathbf{\Sigma}_{12} \\
		\mathbf{\Sigma}_{21} & \mathbf{\Sigma}_{22} \\
	\end{bmatrix}
	)
	\label{eq:pred1}
\end{equation}
with $\mathbf{\Sigma}_{11} \in \mathbb{R}^{m \times m}$, $\mathbf{\Sigma}_{12} \in \mathbb{R}^{m \times n}$, $\mathbf{\Sigma}_{21} \in \mathbb{R}^{n \times m}$, and $\mathbf{\Sigma}_{22} \in \mathbb{R}^{n \times n}$. 

The associated conditional distribution can be represented as

\begin{equation}
	\label{eq:pred2}
	\mathbf{Z}_{1}|\mathbf{Z}_{2} \sim N_{m} 
	(
	\boldsymbol{\mu}_{1}+\mathbf{\Sigma}_{12}\mathbf{\Sigma}_{22}^{-1} (\mathbf{Z}_{2}-\boldsymbol{\mu}_{2})
	,
	\mathbf{\Sigma}_{11}-\mathbf{\Sigma}_{12}\mathbf{\Sigma}_{22}^{-1}
	\mathbf{\Sigma}_{21}
	).
\end{equation}

Assuming that the known measurements vector $\mathbf{Z}_2$ has a zero-mean function (i.e., $\boldsymbol{\mu}_{1} = 0$ and $\boldsymbol{\mu}_{2} = 0)$,  the unknown measurements vector $\mathbf{Z}_1$ can be predicted using~\cite{genton2007separable}

\begin{equation}
	\mathbf{Z}_{1}= \mathbf{\Sigma}_{12} \mathbf{\Sigma}_{22}^{-1} \mathbf{Z}_{2}.
	\label{eq:pred3}
\end{equation}

\section{Contributions}
\label{sec:contrib}
Our contributions can be summarized as follows:
\begin{itemize}
	\item We introduce \emph{ExaGeoStat}, a unified software for computational geostatistics that 
		exploits recent developments in dense linear algebra task-based algorithms associated with dynamic runtime systems.

	\item The \emph{ExaGeoStat} software we propose is able to estimate the statistical model parameters 
		for geostatistics applications and predict missing measurements.
	\item \emph{ExaGeoStat} relies on a single source code to target 
		various hardware resources including shared and distributed-memory systems
		composed of contemporary devices, such as traditional Intel
		multicore processors, Intel manycore processors, and NVIDIA GPU accelerators. 
		This eases the process of software deployment and effectively 
		employs the highly concurrent underlying hardware, thanks to the fine-grained, tile-oriented parallelism
		and dynamic runtime scheduling.

	\item We propose a synthetic dataset generator that can be used to perform
		broader scientific experiments related to computational geostatistics applications.
			
			\item We
		propose an R-wrapper functions for the proposed software (i.e., {\emph{ExaGeoStatR}}) to facilitate the use of our software in the \emph{R} environment~\cite{team2000r}.
		
	\item We evaluate the performance of our proposed software during
		applications using both synthetic and real datasets in terms of
		elapsed time and number of floating-point operations (Gflop/s) on several hardware systems.
	\item We
		assess the quality of the estimation of the Mat\'{e}rn covariance parameters and prediction operation
		achieved by \emph{ExaGeoStat} through a quantitative performance analysis and using both exact and approximation techniques.
\end{itemize}

\section{Climate and Environment Data}
\label{sec:geospat}
In climate and environment studies, numerical models play an important role in
improving our knowledge of the characteristics of the climate system, and of the 
causes of climate variations. These numerical models describe the evolution of 
many variables, for example, temperature, wind speed, precipitation, humidity 
and pressure, by solving a set of equations. The process involves 
physical parameterization, initial condition configuration, 
numerical integration, and data output. In this section, we use the 
proposed methodology to investigate the spatial variability of soil moisture 
data generated by numerical models.
Soil moisture is a key factor in evaluating the state of the hydrological 
process, and has a wide range of applications in weather forecasting, crop yield 
prediction, and early warning of flood and drought. It has been shown that better 
characterization of soil moisture can significantly improve the weather forecasting. 
However, the numerical models often generate very large datasets due to the high 
spatial resolutions, which makes the computation of the widely used Gaussian process 
models infeasible. Consequently, practitioners divide the whole region to 
smaller size of blocks, and fit Gaussian process models independently to each block, 
or reduce the size of the dataset by averaging to a lower spatial resolution.  
However, compared to fitting a consistent Gaussian process model to the entire 
region, it is unclear how much statistical efficiency is lost by such an approximation. 
Since our proposed technique can handle large covariance matrix computations, 
and the parallel implementation of the algorithm significantly reduces the 
computational time, we propose to use exact maximum likelihood inference for 
a set of selected regions in the domain of interest to characterize and 
compare the spatial variabilities of the soil moisture.

We consider high-resolution daily soil moisture data at the top layer of the 
Mississippi River basin, U.S.A., on January 1st, 2004. The spatial resolution is of 
0.0083 degrees, and the distance of one-degree difference in this region is 
approximately 87.5 km. The grid consists of $1830 \times 1329 = 2{,} 432{,} 070$ 
locations with 2,153,888 measurements and 278,182 missing values. We use the same 
model for the mean process as in Huang~\cite{huang2017hierarchical}, and fit a zero-mean Gaussian 
process model with a Mat{\'e}rn covariance function to the residuals; see Huang~\cite{huang2017hierarchical} 
for more details on data description and exploratory data analysis. 


\begin{figure}
	\centering

		\subfigure[LAPACK: Column-major data layout format.]{
			\label{fig:format-cdl}
		\includegraphics[width=0.2\textwidth]{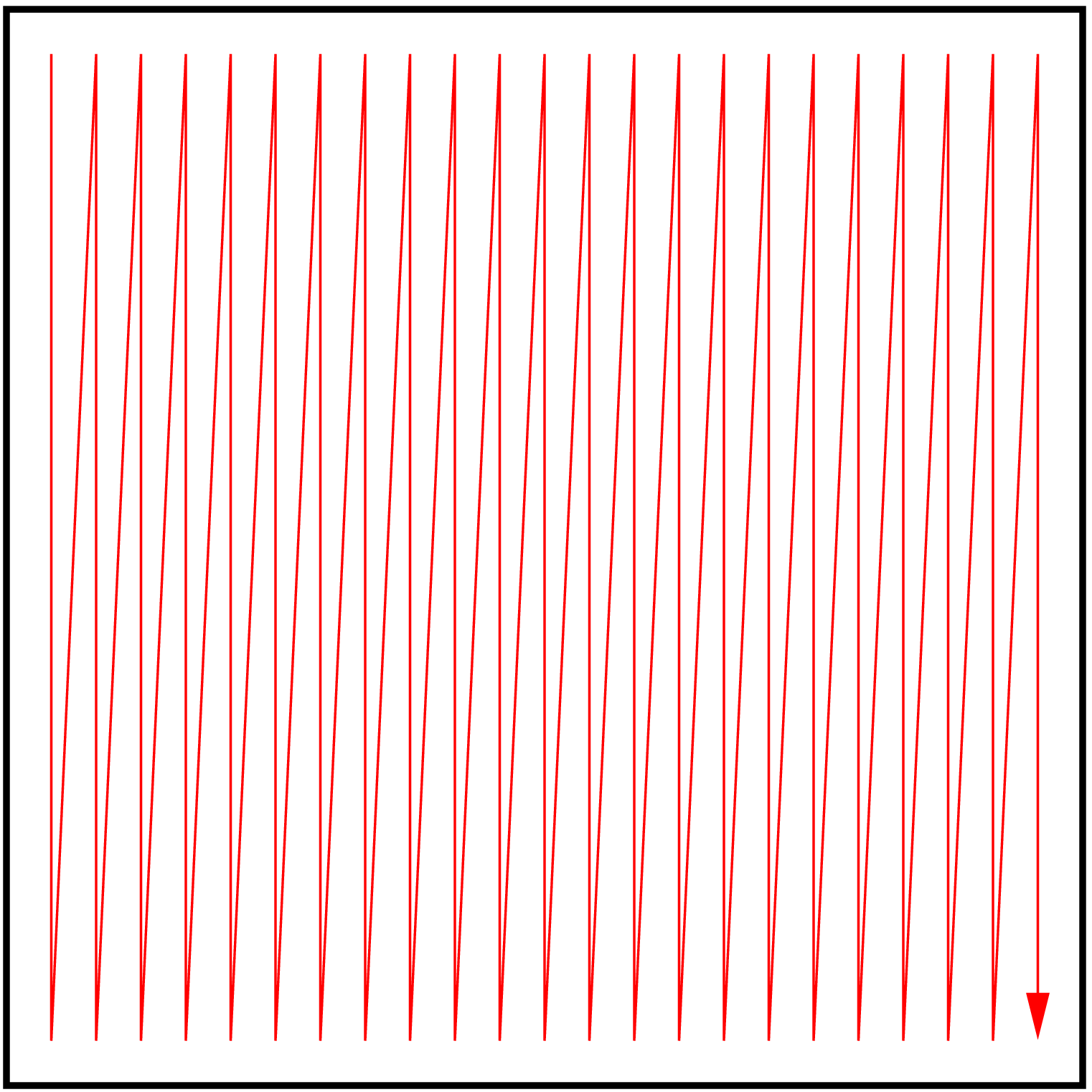}}
		\hspace{2mm}
		\subfigure[Chameleon: Tile data layout format.]{
			\label{fig:format-tdl}
		\includegraphics[width=0.2\textwidth]{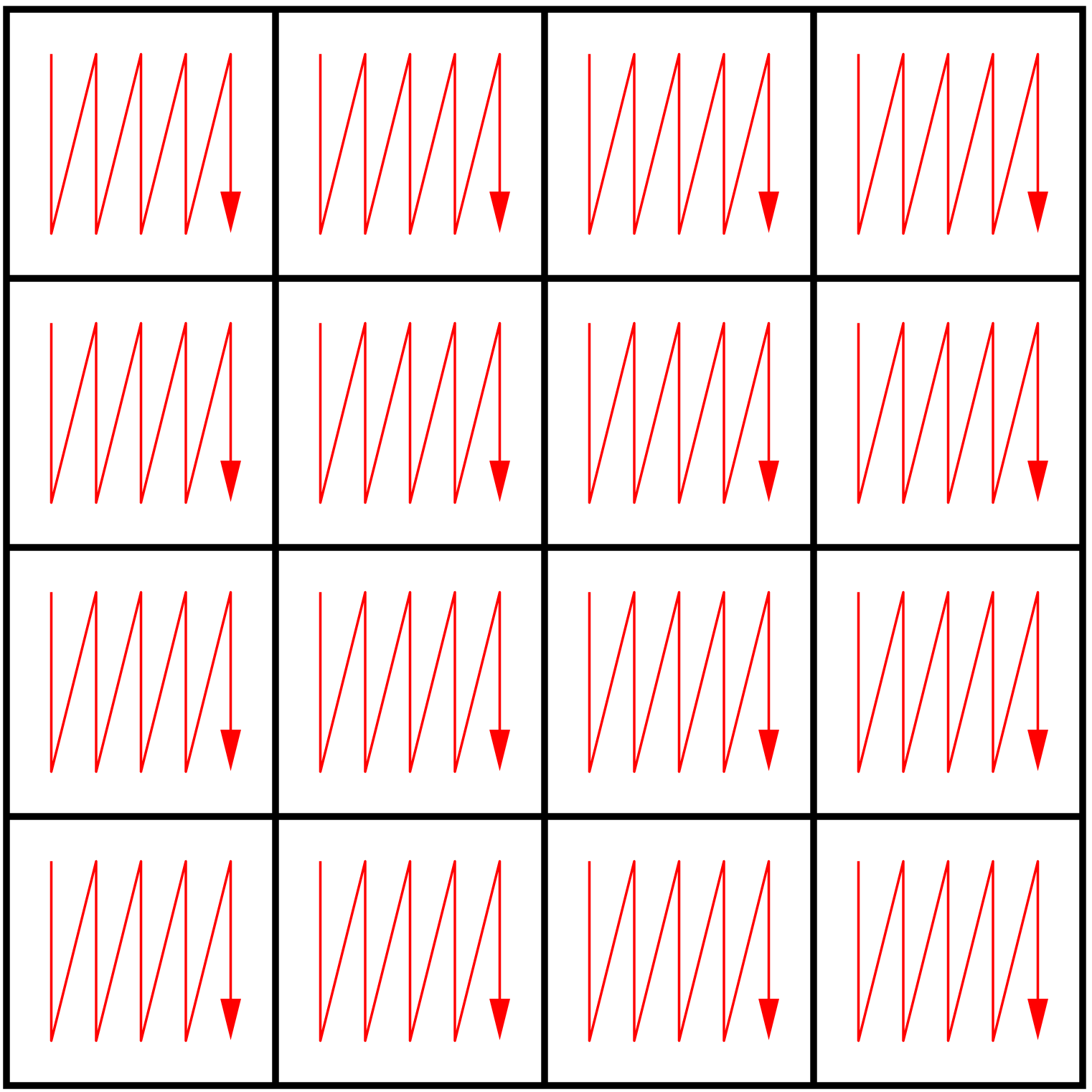}}
	\caption{Data layout format.}
	\label{fig:block-tile-dag}
\end{figure}

\begin{figure}

	\includegraphics[height=0.40\textheight, width=\linewidth]{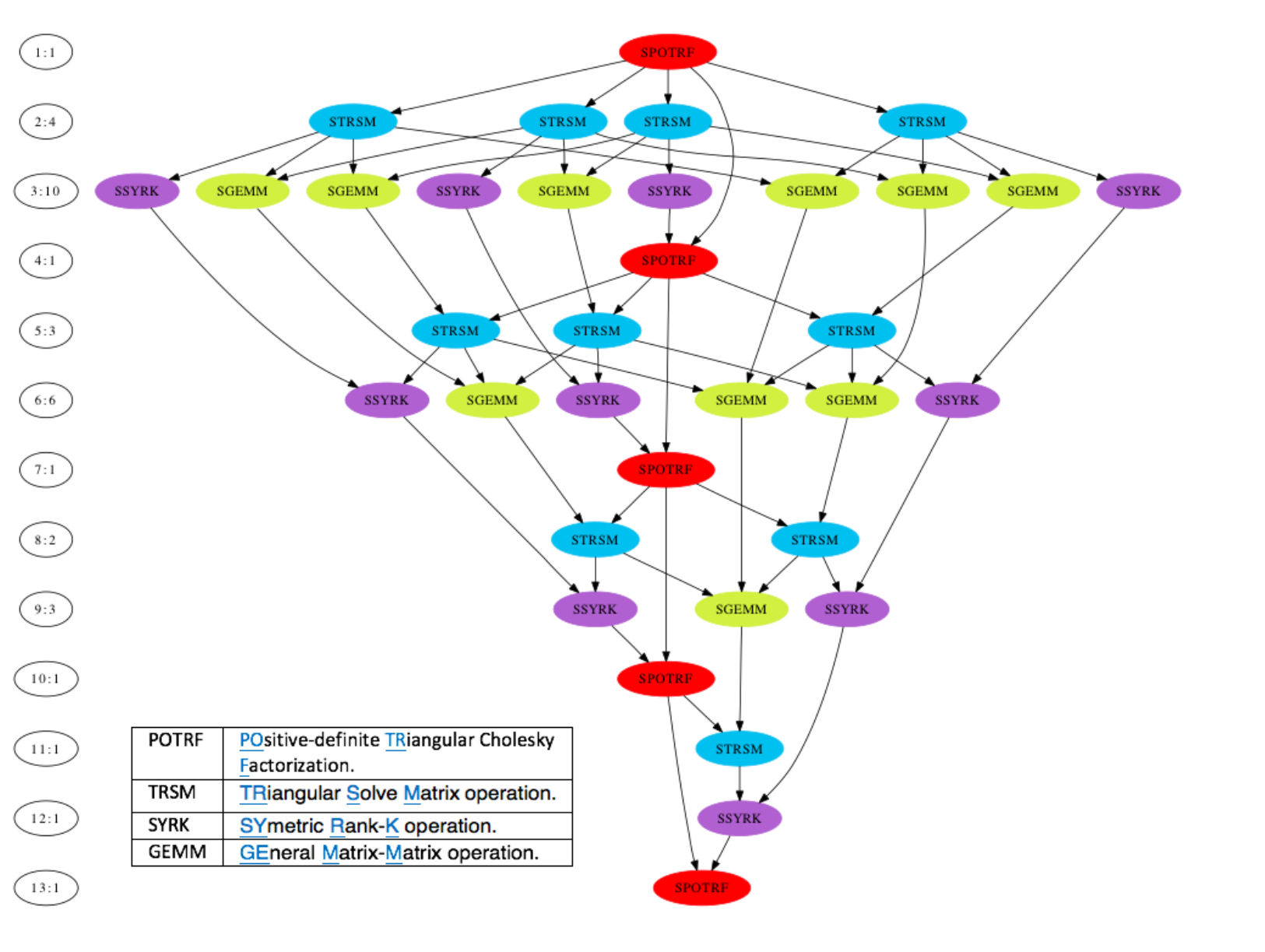}
	\caption{\small Directed Acyclic Graph (DAG) for a Cholesky factorization:
			DAG height corresponds to the length of the critical
	path and the DAG width to the degree of concurrency.}
	\label{fig:dag}
\end{figure}

\section{State-Of-The-Art Dense Linear Algebra Libraries}
\label{sec:sotadla}
This section recalls the latest developments in dense linear
algebra software libraries and their relevant implications.





\subsection{Block Algorithms}

The default paradigm behind LAPACK~\cite{anderson1999lapack}, 
the well-established open-source dense linear algebra library
for shared-memory systems, is block-column algorithms. These algorithms decompose the matrix into successive panel 
and update computational phases, while the matrix is organized in a column-major 
format, see Figure~\ref{fig:format-cdl}. 
The matrix transformations are blocked within the 
panel factorization phase, and applied together at one time during
the update phase. The former is typically memory-bound 
due to the Level-2 BLAS operations, while the latter is compute-intensive
due to the Level-3 BLAS updates occurring on the trailing submatrix. LAPACK 
uses the fork-join paradigm, which has demonstrated scalability issues on multicore
architectures. Its distributed version, ScaLAPACK~\cite{scalapack} follows the same 
paradigm and scatters the matrix
using a two-dimensional block-cyclic data distribution across a grid of processors
to reduce load imbalance and communication overheads.

\subsection{Tile Algorithms}

The tile algorithm methodology~\cite{Agullo_2009_jpcs,chameleon-soft} 
splits the matrix into small tiles instead of tall panels, as
seen in Figure~\ref{fig:format-tdl},
so that updates of the trailing submatrix may be triggered before the current
panel factorization is complete. This fine-grained lookahead method 
exploits more concurrency and enables the maximization of hardware resources by removing
synchronization points between the panel and update computational
phases. The numerical algorithm can then be translated into a Directed
Acyclic Graph (DAG), where the nodes represent tasks and the edges define data dependencies,
as highlighted in Figure~\ref{fig:dag}.

\subsection{Dynamic Runtime Systems}

Once the tasks are defined with their respective data
dependencies, a dynamic runtime system~\cite{augonnet2011starpu,duran2009proposal, EdwardsTS14} 
may be employed directly on the sequential code to schedule the
various tasks across the underlying hardware resources. Its role
is to ensure that the data dependencies are not violated. These runtimes enhance the software productivity by abstracting
the hardware complexity from the end users. They are also capable of reducing
load imbalance, mitigating data movement overhead, and increasing occupancy
on the hardware.

\section{The ExaGeoStat Software}
\label{sec:impl}
\subsection{General Description}
We propose a unified computational software for geostatistical
climate and environmental applications based on the
maximum likelihood approach.
Since the covariance matrix is symmetric and
positive-definite, the computation of the maximum likelihood
consists of the Cholesky factorization and its corresponding
solver which uses measurements vector $\mathbf{Z}$ as the right-hand side.
The log-determinant is calculated from the Cholesky factor
simply by computing the product of the diagonal entries.

The objective of this software is not only
to solve the maximum likelihood problem for a given
set of real measurements, $\mathbf{Z}$, on $n$
geographic locations, but also to predict a set of unknown measurements at new locations. The proposed 
software also provides a generic tool for generating a reference set of synthetic
measurements and locations for statisticians, which generates test cases
of prescribed size for standardizing comparisons with other methods.


Our proposed software has two different execution modes for dealing with synthetic
and real datasets. In {\em testing mode}, \emph{ExaGeoStat} generates the measurements data based on a given vector
$\boldsymbol \theta$ = $(\theta_1,\theta_2,\theta_3)^\top$, where $\theta_1$ is
the variance parameter, $\theta_2$ is the range parameter, and $\theta_3$ is the
smoothness parameter. In this case, the resulting $\boldsymbol {\reallywidehat \theta}$ vector, which maximizes the likelihood function,
should contain a set of values close to the initial $\boldsymbol \theta$ vector. Moreover, testing the prediction accuracy can
be done by choosing random measurements from the given synthetic dataset and use the generated model to predict these
measurements using the other known measurements. The accuracy of the predictions can be verified by comparing random
measurements from the given synthetic dataset to the corresponding generated measurements from the model.

In {\em application mode}, both the measurements and the locations data are given,
so the software is only used to evaluate the MLE function by estimating the parameter 
vector, $\boldsymbol {\reallywidehat \theta}$. The generated model can be used to predict unknown
measurements at a set of new locations.

\subsection{Software Infrastructure}

\emph{ExaGeoStat} internally relies on Chameleon, a high performance
numerical library~\cite{chameleon-soft}. Based on a tile algorithm,
Chameleon is a dense linear algebra library that provides high-performance
solvers. Chameleon handles dense linear algebra operations through a sequential
task-based algorithms. It features a backend with links to several
runtime systems, and in particular, the StarPU dynamic runtime system,
which is preferred for its wide hardware architecture
support (Intel manycore, NVIDIA GPU, and distributed-memory systems).

StarPU deals with the execution of generic task graphs,
which are generated by a sequential task flow (STF) programming model.
The tasks are sequentially given to StarPU with hints of the data
dependencies (e.g., read, write, and read-write). The StarPU runtime schedules
the given tasks based on these hints. The main advantage
of using a runtime system that relies on task-based implementations such as StarPU 
is to become oblivious of the targeted hardware architecture. This kind of
abstraction improves both the user productivity and creativity.
Multiple implementations of the same StarPU tasks are generated for:
CPU, CUDA, OpenCL, OpenMP, MPI, etc. At runtime,  StarPU decides
automatically which implementation will achieve the highest performance.
For the first execution, StarPU generates a set of cost models that
determine best hardware for optimal performance during the given tasks. This set of cost models may be saved for future executions.

\begin{figure}
	\centering
	\includegraphics[width=5.5 cm]{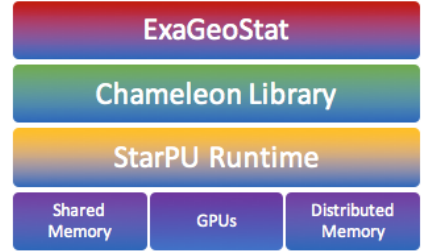}
	\caption{\emph{ExaGeoStat}  software.}
	\label{fig:software}
\end{figure}

Figure~\ref{fig:software} shows the structure of the \emph{ExaGeoStat} software. 
It has three main layers: \emph{ExaGeoStat}, which includes the
upper-level functions of the software; the Chameleon library, which provides solvers
for the linear algebra operations; and the StarPU runtime,  which translates the software
for execution on the appropriate underlying hardware.

\subsection{The Optimization Framework}
Finding the parameter vector ${\boldsymbol {\reallywidehat \theta}} =(\theta_1,\ldots,\theta_q)^\top$, that
maximizes the likelihood function requires several iterations  of the log-likelihood
evaluation. In our proposed software, we rely on an open-source C/C++ nonlinear
optimization toolbox, NLopt~\cite{johnson2014nlopt}, to perform the optimization task.
The NLopt package contains $20$ global and local optimization algorithms. NLopt solves nonlinear optimization 
problems of the form $\min_{\mathbf{x} \in
\mathbb{R}^{q}} f(\mathbf{x})$, where $f$ represents the objective function and $\mathbf{x}$ represents
the $q$ optimization parameters, i.e., the parameter vector. Because we are targeting
a nonlinear problem with a global maximum point, we selected BOBYQA for our proposed platform.

BOBYQA is one of the optimization algorithms available in the sequential Nlopt package
to optimize the MLE function.
It is a numeric, global, derivative-free and bound-constrained optimization algorithm. It generates
a new computed point on each iteration by solving a {\em trust region} subproblem
subject to given constraints~\cite{powell2009bobyqa}, in our case, only upper and lower bound constraints are
used.  Though BOBYQA does not require the evaluation of the derivatives
of the cost function, it employs an iteratively updated quadratic model of the objective, so there
is an implicit assumption of smoothness.

The master process feeds the optimization
 black box BOBYQA function with the current ${\boldsymbol \theta}$ vector, which produces the 
 resulting likelihood value. This likelihood value gets broadcasted to all other running processes,
 which, in return, carry on with subsequent computations. This optimization step 
 is then repeated with a new parameter vector ${\boldsymbol \theta}$ at each iteration,
 until convergence is reached.
 

As with the linear algebra software, we employ these optimization frameworks without novel contributions herein, 
in order to achieve the practical synthesis of well-understood components.
For now, we merely design the interfaces of these codes, which, in the case of BOBYQA, consists mainly
of callbacks to the log-likelihood function with a sequence of Mat\'ern triples that must be evaluated using the
measurement vector and the covariance matrix.  The log-likelihood function may need to be evaluated 
many times, but after the initial factorization the cost of each estimation step should remain constant.

\subsection{Synthetic Data Generator}
\emph{ExaGeoStat}  provides an internal data generator that is used here to demonstrate
the accuracy of the software. This data generator can also be used as a stand-alone tool
to generate sets of guided synthetic data for experiments with specific needs or conditions.

Given $n$ locations that are uniformly but randomly distributed,
the covariance matrix $\mathbf{\Sigma}$
can be built using the Mat\'ern covariance function (i.e., Equation (2)).
This covariance matrix can be used to
generate a measurement vector $\mathbf{Z}$ from normal variates
at the generated $n$ locations, as follows

$\\
{\boldsymbol \Sigma}={\bf L} \cdot {\bf L}^\top \quad \Rightarrow$ Cholesky factorization.\\
$\mathbf{Z}={\bf L} \cdot \mathbf{e} \qquad \Rightarrow$ where $ e_i  \sim N(0,1). \\ $

The data generator
tool is shown in Algorithm~\ref{alg:SDGA}. To generate a synthetic measurement vector
$\bf {Z}$, the algorithm randomly generates a set of $n$ locations (line 2). Then, the distance matrix $\bf D$ is generated between
these $n$ random locations (line 3). In line 4, an initial covariance matrix $\boldsymbol \Sigma$ is generated using
the $\bf D$ matrix and the initial parameter vector ${\boldsymbol \theta}$. In line 5, a Cholesky factorization step
is performed on the covariance matrix $\mathbf{\Sigma}$ by using the Chameleon routine {\bf dpotrf} to generate the lower triangular
matrix $\mathbf{L}$. After generating the initial normal random vector, $\mathbf{e}$, a single matrix-vector multiplication operation is 
performed using the lower triangular matrix $\mathbf{L}$ and the random vector $\mathbf{e}$ to initiate the synthetic measurement 
vector $\mathbf{Z}$ (lines 6-7). Here, the Chameleon routine {\bf dtrmm}  is used.
\begin{algorithm} [htp]
\setstretch{0.9}
	\caption{\bf: Synthetic Data Generator Algorithm }
	\label{alg:SDGA}
	\begin{algorithmic}[1]
		\State Input: initial parameter vector ${\boldsymbol \theta}$
		\State  Uniform random generation of $n$ locations 
		\State  $\mathbf{D}$ = genDistanceMatrix ($n$, $n$)
		\State  $\mathbf{\Sigma}$ = genCovMatrix ($\mathbf{D}$, ${\boldsymbol \theta}$)
		\State  $\mathbf{L}{\bf L}^{\top}$ = dpotrf ($\mathbf{\Sigma}$)   $\Rightarrow$ 
		Cholesky factorization  ${\boldsymbol \Sigma}={\mathbf L}{\mathbf L}^{\top}$
		\State  Normal random generation of a vector $\mathbf{e}$
		\State  ${\bf Z}$ = dtrmm (${\mathbf L}, \mathbf{e}$)    $ \Rightarrow$ Solve ${\mathbf Z} = {\mathbf L}*\mathbf{e}$
	\end{algorithmic}
\end{algorithm}

\subsection{Likelihood Evaluation}
As mentioned, our software has two different running modes: 
{\em testing mode} to build a statistical model based on a given set of parameters with the aid of a synthetic 
set of data (i.e., measurements and locations) and {\em application mode} where measurements and locations
data are given to estimate the statistical model's parameters for future prediction of unknown measurements at a new set of locations.

For both modes, with a given measurement vector $\mathbf{Z}$ and distance matrix $\mathbf{D}$,
the likelihood function can be evaluated using a set of routines from the Chameleon
library. The evaluation algorithm based on Equation (1) is presented in detail in  Algorithm~\ref{alg:MLE}. 
The inputs to the evaluation algorithm are the measurement vector $\mathbf{Z}$,
distance  matrix $\mathbf{D}$, and  parameter vector ${\boldsymbol \theta}$ (line 1).
The algorithm generates the covariance matrix $\boldsymbol \Sigma$ (line 2) 
using the Mat{\'e}rn function given by Equation (2).
In line 3, a Cholesky factorization step is performed on the covariance matrix $\boldsymbol \Sigma$ by using the 
{\bf dpotrf} routine to generate the lower triangular matrix $\mathbf{L}$. In line 4, a triangular solver {\bf dtrsm} is used to solve
$\mathbf{L} \times \mathbf{Z_{new}}=\mathbf{Z_{old}}$. Both the log-determinant and dot product operations are performed
in lines 5-6.  In line 7, the likelihood value $\ell$, which should be maximized, is calculated based on
the $dotscalar$ and $logscalar$ values.

To find the maximum likelihood value, this algorithm is called several times with different parameter vectors $\boldsymbol{\theta}$ with
the help of the used optimization function.

\begin{algorithm} [htp]
\setstretch{0.9}
	\caption{\bf: Log-likelihood Evaluation Algorithm}
	\label{alg:MLE}
	\begin{algorithmic}[1]
		\State Input:  measurement vector  $\bf {Z}$, distance matrix $\bf D$, and initial parameter vector ${\boldsymbol \theta}$
		\State  $\bf {\Sigma}$ = genCovMatrix ($\bf{Z}$, $\bf D$, ${\boldsymbol \theta}$)
		\State ${\bf L}{\bf L}^{\top}$ = dpotrf ($\boldsymbol \Sigma$)   $ \Rightarrow$  Cholesky factorization  ${\boldsymbol \Sigma}={\bf L} \times {\bf L}^{\top}$
		\State  ${\bf Z_{new}}$ = dtrsm (${\bf L},\bf{Z_{old}}$)   $  \Rightarrow$ Triangular solve ${\boldsymbol \Sigma}*{\bf Z_{new}} = {\bf Z_{old}}$
		\State  $logscalar$ = computeLogDet ($\boldsymbol \Sigma$)  $  \Rightarrow$  The log determinant $\log |{\boldsymbol \Sigma}|$
		\State  $dotscalar$ = computeDotProduct ($\bf{Z}$, $\bf{Z}$)   $  \Rightarrow$ The dot product of $\bf{Z} \times \bf{Z}$
		\State  ${\ell}$= $-0.5 \times dotscalar -0.5 \times logscalar - (\frac{n} {2})  \log(2\pi)$
	\end{algorithmic}
\end{algorithm}

The main goal of Algorithm~\ref{alg:MLE} is to calculate the likelihood function using a certain
${\boldsymbol \theta}$ vector. However, our statistical model relies on finding the parameter vector ${\boldsymbol {\reallywidehat \theta}}$, which
maximizes the value of the likelihood function $\ell$. Thus, BOBYQA optimization algorithm
is used with the ${\boldsymbol \theta}$ vector and the $\ell$ value to find the optimized vector ${\boldsymbol {\reallywidehat \theta}}$ 
for the given problem ($\bf{Z}$, $\mathbf{\Sigma}$). It is difficult to determine in advance the average number of 
iterations needed to maximize the likelihood function because it depends on several factors, such as  the optimization 
algorithm, the initial parameters ${\boldsymbol \theta}$, and the maximum acceptable relative tolerance (i.e.,
the measure of error between the current solution and the previous solution).

\subsection{Prediction}
In the likelihood estimation step, we aim to construct a statistical model based on
 estimated parameters (i.e., ${\boldsymbol {\reallywidehat \theta}}$ vector). 
This model can be used for predicting $m$ unknown measurement in the vector 
${\mathbf Z}_{1}$  with the aid of $n$ known measurement in the vector ${\mathbf Z}_{2}$ 
(see Equation (5)). The prediction operation can also be implemented  using a set of
 routines from the Chameleon library.

Algorithm~\ref{alg:pred} shows the prediction algorithm in details. The algorithm has a set of inputs:
 the parameter vector $\bf{\reallywidehat \theta}$, 
the measurement vector ${\bf Z}_{2}$, a vector of the known $n$ locations, 
and a vector of the new $m$ locations with unknown measurement vector
${\bf Z}_{1}$ (line 1). The algorithm aims to predict the measurement vector ${\bf Z}_{1}$ 
 at the given $m$ locations (line 2). In lines 3 and 4, two distance
matrices are generated: ${\bf D}_{22}$ between two sets of the observed $n$ locations, 
and ${\bf D}_{12}$ between 
the given unobserved $m$ locations and the observed $n$ locations. These distance matrices are used 
to construct  two covariance matrices, $\bf {\Sigma_{22}}$ and
${\bf \Sigma}_{12}$ (lines 5-6). In line 7, the {\bf dposv} routine is used to solve the system of
 linear equation  $ {\bf Z} \times {\bf X} =  {\bf \Sigma}_{22}$. 
In line 8, the unknown measurement vector, ${\bf Z}_{1}$, can be calculated using 
the {\bf dgemm} routine (i.e., matrix-matrix multiplication),
${\bf Z}_{1}= {\bf \Sigma}_{12} \times {\bf X}$.

\begin{algorithm} [htp]
\setstretch{0.9}
	\caption{\bf: Prediction Algorithm}
	\label{alg:pred}
	\begin{algorithmic}[1]
		\State Input:  parameter vector ${\boldsymbol {\reallywidehat \theta}}$, known measurements vector ${\bf Z}_{2}$, observed $n$ locations, and new $m$ locations.
		\State Output: unknown measurements vector ${\bf Z}_{1}$
		\State  ${\bf D}_{22}$= genDistanceMatrix ($n$, $n$)
		\State  ${\bf D}_{12}$= genDistanceMatrix ($m$, $n$)
		\State   ${\bf \Sigma}_{22}$= genCovMatrix (${\bf D}_{22}$, ${\boldsymbol {\reallywidehat \theta}}$)
		\State   ${\bf \Sigma}_{12}$= genCovMatrix (${\bf D}_{12}$, ${\boldsymbol {\reallywidehat \theta}}$)
		\State  $\bf X $ = dposv (${\bf \Sigma}_{22}$, ${\bf Z}_{2}$)        $  \Rightarrow$     Compute the solution to a system of linear equation $ {\bf Z} \times {\bf X} =  {\bf \Sigma}_{22}$
		\State  ${\bf Z}_{1}$  = dgemm (${\bf \Sigma}_{12}$, ${\boldsymbol X}$)        $  \Rightarrow$     Performs the matrix-matrix operation ${\bf Z}_{1}= {\bf \Sigma}_{12} \times {\bf X}$
	\end{algorithmic}
\end{algorithm}

\subsection{Facilitating \emph{ExaGeoStat} Adoption for Statisticians}
Statisticians rely heavily on \emph{R}, a high productivity simulation environment, to rapidly
deploy and assess their algorithms, especially when applied to big data problems, as in 
climate and environmental research studies.
 Therefore, we provide \emph{R}-wrappers to our main computational functions through a separate package called 
 \emph{ExaGeoStatR}. These R functions should help in disseminating our software
 toward a large computational statistician community.
 To the best of our knowledge, most of existing \emph{R} solutions for the MLE
   problem are sequential and restricted to limited data sizes such as 
   {\em{fields}} package provided by the University Corporation
    for Atmospheric Research (UCAR)~\cite{ucar_fileds}.

\subsection{Independent Blocks (IND) Approximation}
Approximation means exist to reduce the algorithmic complexity when dealing with very large and irregularly spaced geospatial data. 
Several previous studies show how the likelihood estimation problem can be adapted to provide different methods
 of hierarchical low-rank approximations. These methods have shown their effectiveness in both computation and
  accuracy~\cite{datta2016hierarchical, huang2017hierarchical, 2017arXiv170904419L, nychka2015multiresolution, stein2013statistical, sun2016statistically}.

As mentioned in the introduction section, this paper is mostly focusing on the exact computation of large geospatial data.
The quality of exact computation may be demonstrated by comparing it with traditional approximation approaches on the same problem.
Thus, in this section, we highlight one of commonly used approximation strategies, i.e., Independent blocks (IND).
Such a strategy has been used in different studies and turns out to be a suitable
way to reduce the complexity of evaluating the likelihood function on large-scale datasets~\cite{huang2017hierarchical, stein2013statistical}. 

\begin{figure}
	\centering
	\includegraphics[width=4 cm, height= 4 cm ]{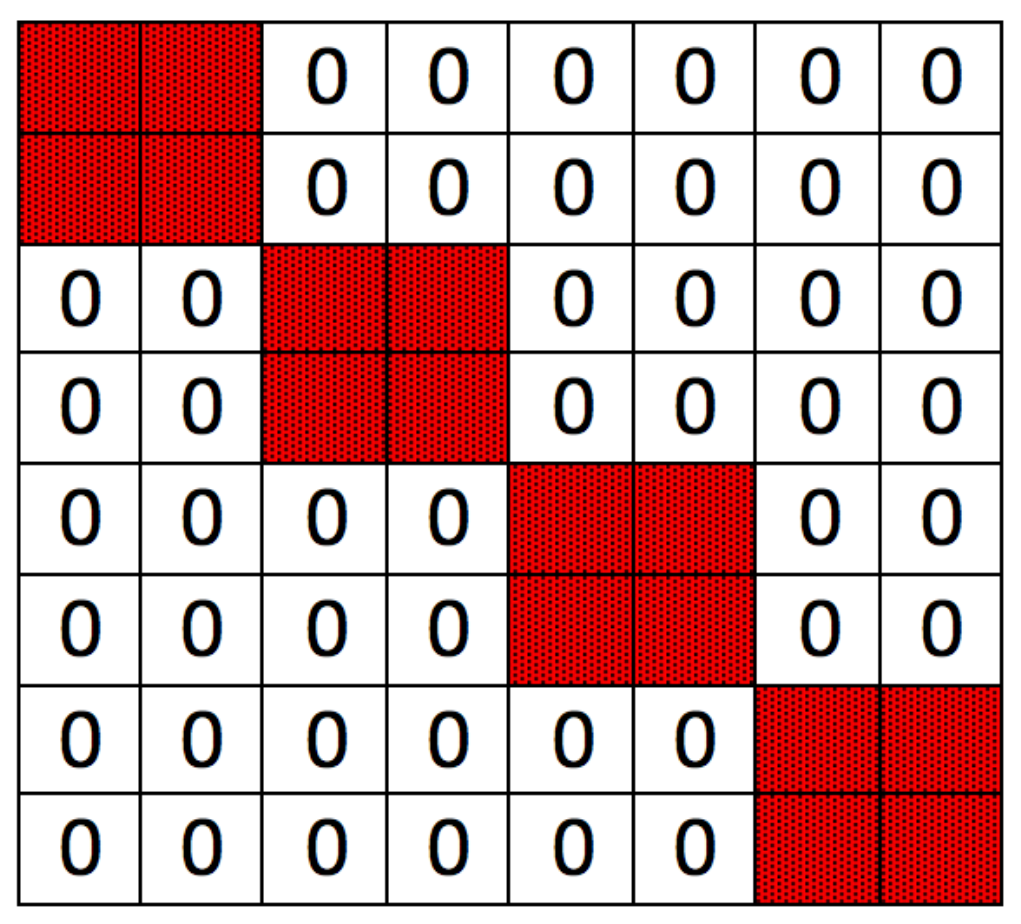}
	\caption{An example of Independent Blocks (IND) approximation technique on a diagonal 2-by-2 super tile matrix. }
	\label{fig:diag-approx}
\end{figure}

The IND approximation technique proceeds by annihilating off-diagonal tiles, since their contributions 
as well as their qualitative impact on the overall statistical problem may be limited. The implementation of this
approximation technique maps well with the inherent tile algorithms of \emph{ExaGeoStat} and enables to expose
tuning parameters, which trades off performance and statistical efficiency. One of these tunable parameters
is the size of the diagonal super tile, which determines how many tiles, around the diagonal, need to be aggregated together.
A large diagonal super tile basically integrates more statistical contributions from the original problem at the expense of
performing more operations.
Figure~\ref{fig:diag-approx} shows an example of the IND approximation technique using a diagonal 2-by-2 super tile matrix.


The IND approximation technique generates a matrix where elements from off-diagonal 
tiles are set to zero. In this case, applying Cholesky factorization
to the whole matrix is unnecessary and time-consuming. Thus, we propose a 
modified version of the well-known tile Cholesky factorization
algorithm presented in~\cite{Agullo_2009_jpcs}. The modified version 
is aware of the new sparse structure of the matrix 
and avoids zeros-tiles during the computation, which speeds-up the Cholesky 
factorization operation for the whole matrix.






\section{Experimental Results}
\label{sec:perf}
\subsection{Environment Settings}
We evaluate the performance of the proposed software on a wide range of 
manycore-based systems: a dual-socket 28-core Intel Skylake Intel Xeon Platinum
8176 CPU running at 2.10 GHz, a dual-socket 18-core Intel Haswell 
Intel Xeon CPU E5-2698 v3 running at 2.30 GHz and equipped with 8 NVIDIA K80s 
(2 GPUs per board),  a dual-socket 14-core Intel Broadwell 
Intel Xeon E5-2680 V4 running at 2.4 GHz,  Intel manycore Knights Landing (KNL) 7210 
chips with 64 cores, a dual-socket 8-core Intel Sandy Bridge Intel Xeon CPU E5-2650 running at 2.00 GHz,
and a dual-socket Intel IvyBridge Intel Xeon CPU E5-2680 running at 2.80 GHz.

For the distributed memory experiments, we use KAUST's 
Cray XC40 system, Shaheen, with 6,174 dual-socket compute nodes
based on 16-core Intel Haswell processors running at
2.3 GHz, where each node has 128 GB of DDR4 memory. 
The Shaheen system has a total of 197,568
processor cores and 790 TB of aggregate memory.

Our software is compiled with gcc v4.8 and linked against 
the Chameleon library v0.9.1 with HWLOC v1.11.5, StarPU v1.2.1, Intel MKL v11.3.1, and NLopt v2.4.2 optimization
libraries. The LAPACK implementation is the multithreaded version from the vendor optimized Intel MKL v11.3.1
numerical library, available on each platform.

\begin{figure}
	\centering
	\includegraphics[width=4.5 cm, height= 4.5 cm ]{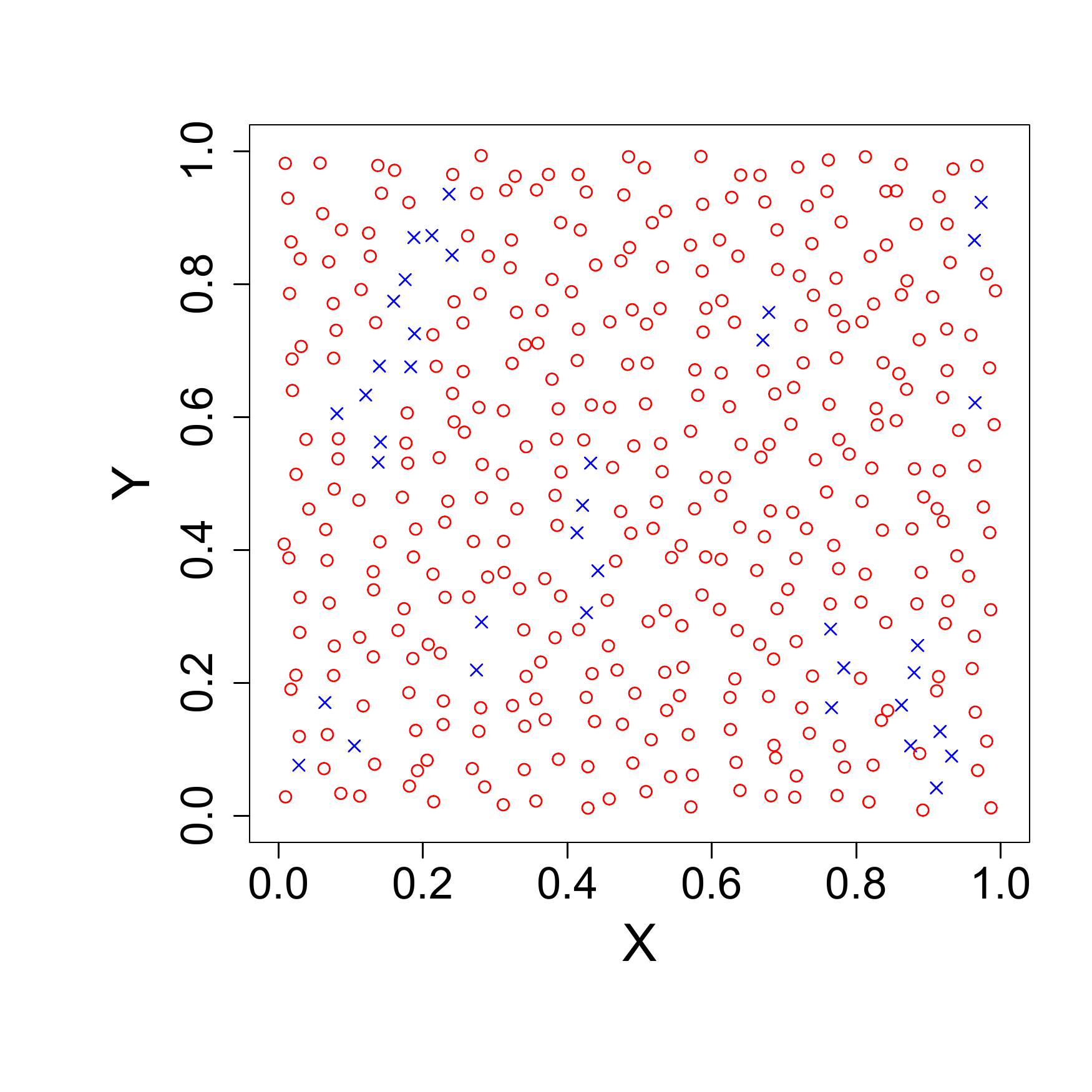}
	\caption{An example of  400 points irregularly distributed in space, with 362 points (\textcolor{red}{$\circ$}) for maximum likelihood estimation and 38 points (\textcolor{blue}{$\times$}) for prediction validation.}
	\label{fig:irregular-space}
\end{figure}

In this study, the synthetic datasets are generated at irregular locations in a two-dimensional space
with an unstructured covariance matrix~\cite{huang2017hierarchical,sun2016statistically}. 
To ensure that no two locations are too close, the data locations are generated using $n^{1/2}(r-0.5+X_{rl},l-0.5+Y_{rl})$ for 
$r,l \in \{1,\dots, n^{1/2}\}$, where $n$ represents the number of locations, and $X_{rl}$ and $Y_{rl}$ are generated using uniform distribution on ($-$0.4, 0.4).  
Figure~\ref{fig:irregular-space} shows a drawable example of 400 irregularly spaced grid locations in a square region. We only use such a small example to highlight
our methodology to generate geospatial data, however,  this work uses synthetic datasets up to $49 \times 10^{10}$ locations (i.e., 700 K $\times$ 700 K).

\subsection{Quantitative Results Assessment}
\subsubsection{Likelihood Estimation Performance}

\begin{figure}[!ht]
	\centering
	\subfigure[Intel two-socket Haswell.]{
		\label{fig:time-haswell}
		\includegraphics[width=0.9\linewidth]{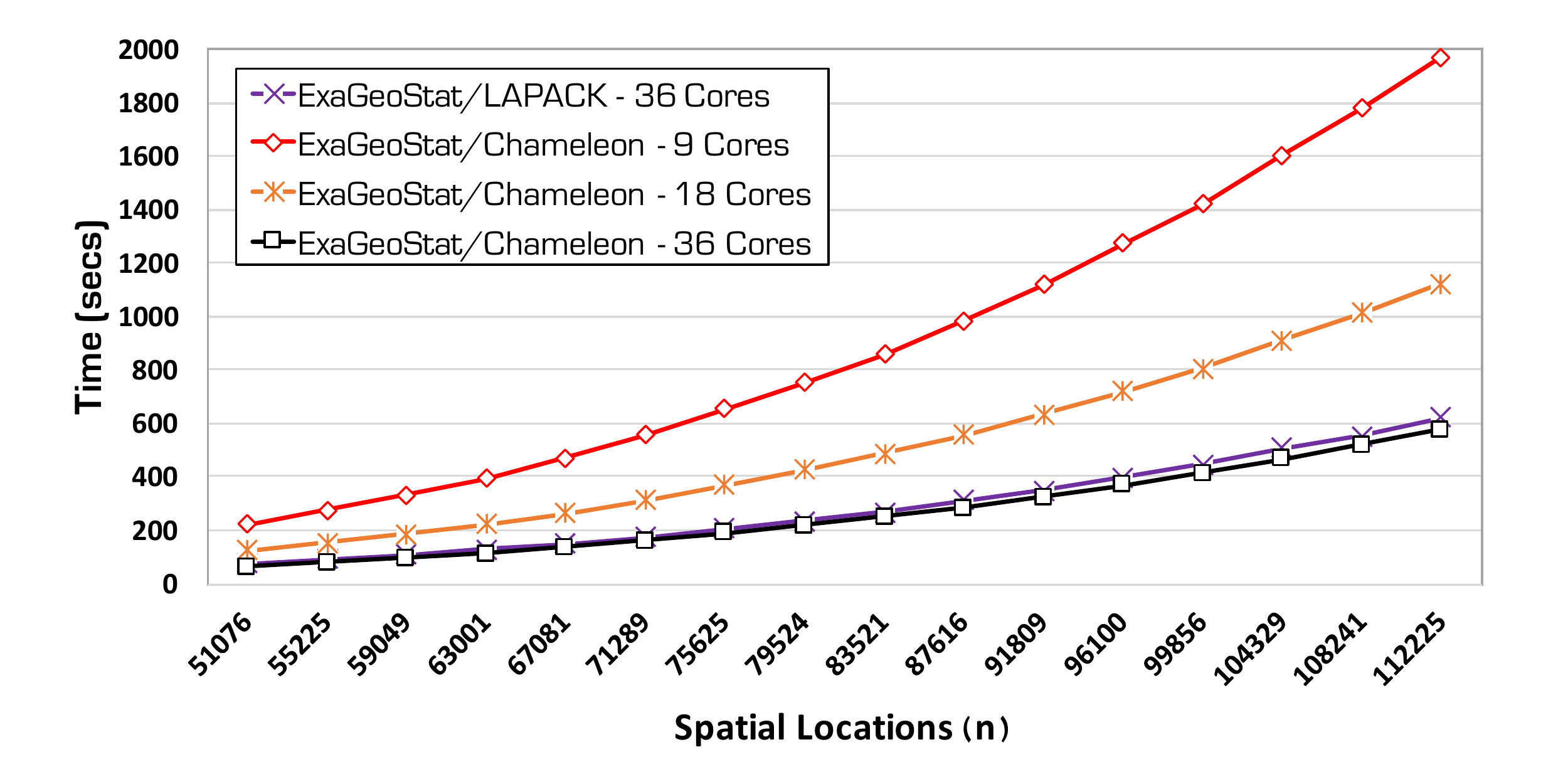}
	}
	\subfigure[Intel two-socket Broadwell.]{
		\label{fig:time-broadwell}
	\includegraphics[width=0.9\linewidth]{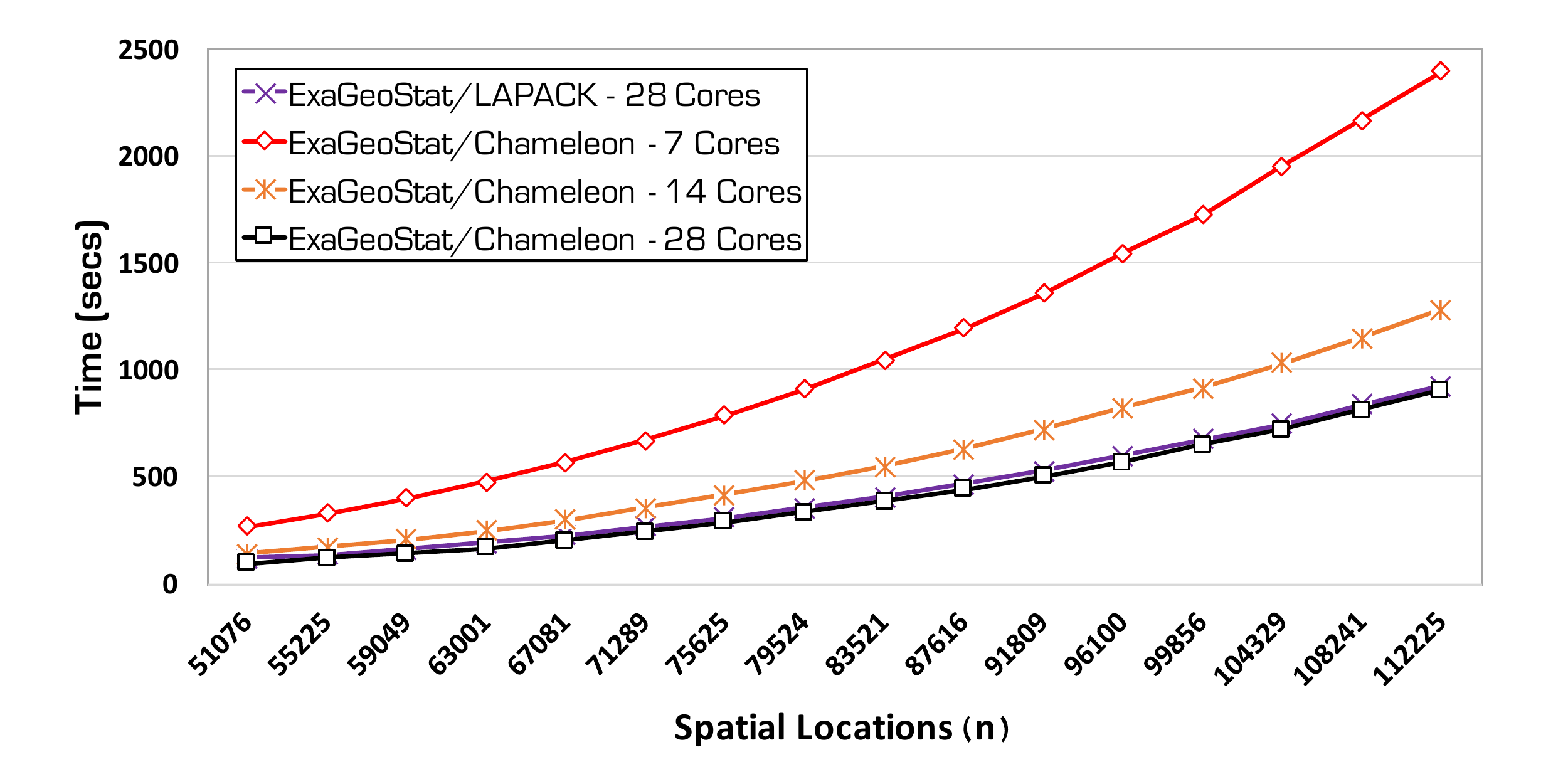}}
	\subfigure[Intel two-socket Knights Landing (KNL).]{
		\label{fig:time-knl}
	\includegraphics[width=0.9\linewidth]{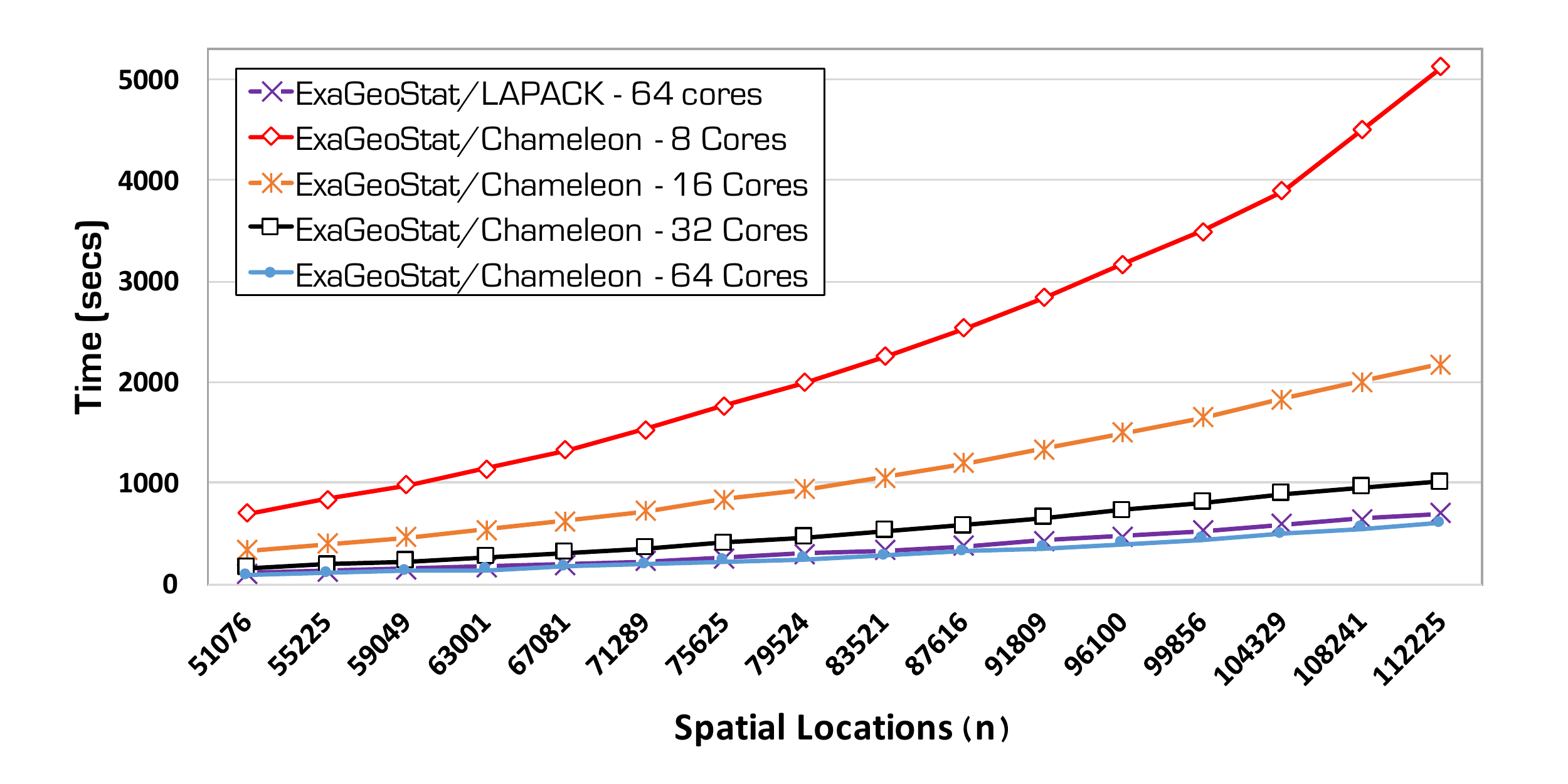}}
	\subfigure[Intel two-socket Haswell + NVIDIA K80.]{
		\label{fig:time-falcon}
	\includegraphics[width=0.9\linewidth]{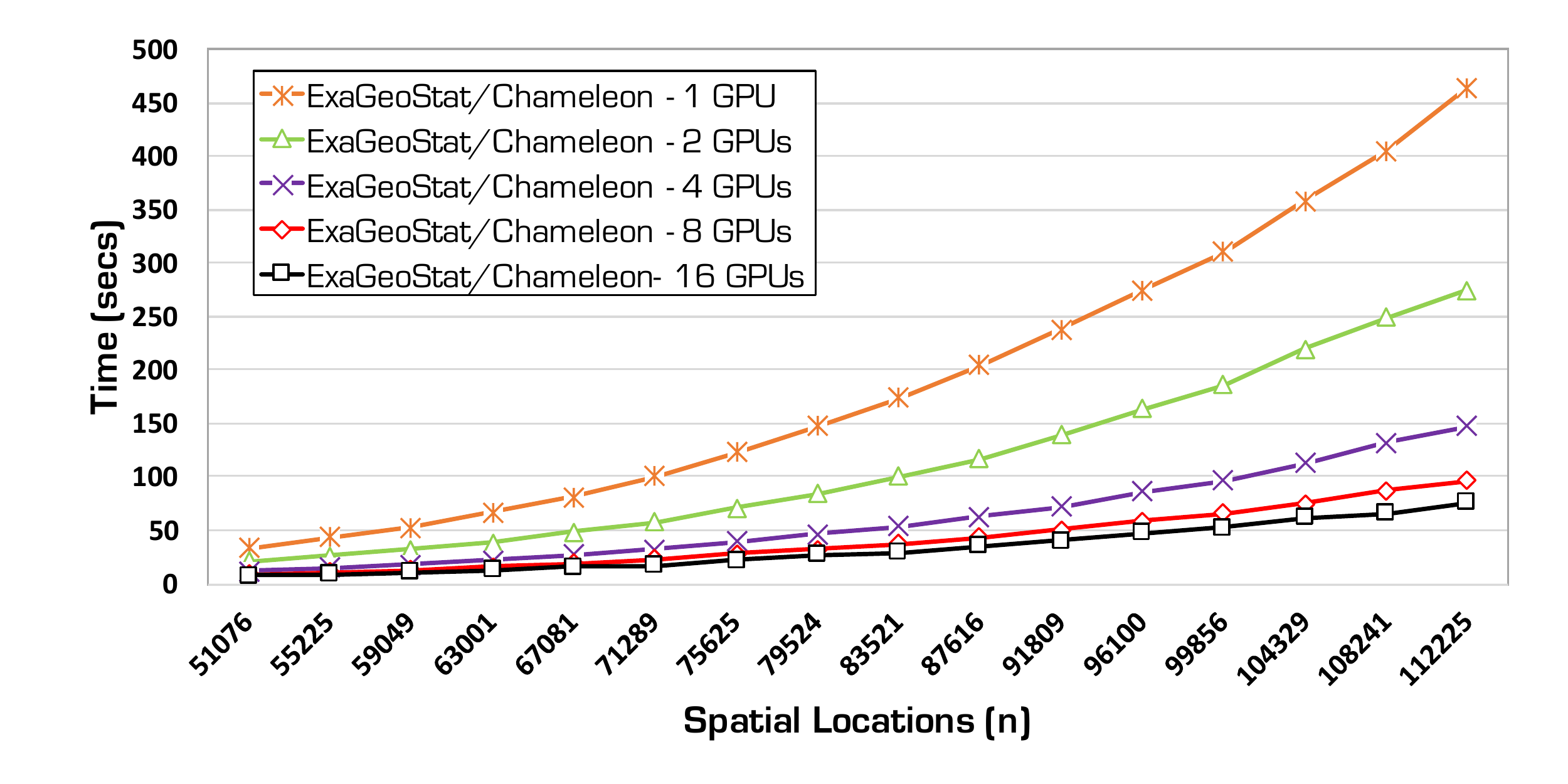}}
	
	\caption{Time for one iteration of the likelihood estimation.}
	\label{fig:execution-time}
\end{figure}
The first set of experiments highlights the execution time for a single iteration of
the MLE algorithm on different target systems. 
We compare our software with the numerical library 
LAPACK~\cite{anderson1999lapack}. LAPACK is considered  the main 
backbone of existing MLE implementations for geostatistical applications~\cite{gil2013codonphyml}.

We report the results of  several experiments on different hardware
architectures: shared-memory, GPUs, and distributed-memory. As the maximum likelihood
estimation problem includes an optimization operation with several likelihood estimation
iterations, we report only the time to finish one iteration of the  likelihood estimation. The LAPACK curves
represent the performance of the LAPACK-based \emph{ExaGeoStat} using all threads. The figures show that
the Chameleon-based \emph{ExaGeoStat} outperforms the LAPACK variant on different platforms when using the same number of threads.
The figures also display the scalability of \emph{ExaGeoStat} when using different numbers of threads.

Figure~\ref{fig:time-haswell} shows the execution  time for a single iteration of  MLE with 7, 
18, and 36 threads compared to
the LAPACK implementation on a Haswell processor. As shown, our implementation achieved a 1.14$\times$ speedup compared with the
LAPACK implementation by exploiting up to 70\% of the peak performance of the Haswell processor.

Figure~\ref{fig:time-broadwell} shows the execution time with 7, 14, and 28 threads compared to
the LAPACK implementation on a Broadwell processor. Our chameleon-based platform speeds up the
execution time by 1.25$\times$ with 28 threads compared to the LAPACK implementation. 
Moreover, our implementation with 28 threads is able to reach over 53\% of 
the total peak performance of the Broadwell processor, while the LAPACK implementation
can only reach 47\% of the peak performance. The figure also shows
scalability using different numbers of threads.

The performance of our proposed platform running
on an Intel Knights Landing (KNL) processor is reported in Figure~\ref{fig:time-knl}.
The platform is easily scaled to accommodate  different numbers of threads (i.e., 4, 8, 16, 32, and
64). 
Using the entire capability of KNL -- 64 threads --
we achieve an overall speedup of 1.20$\times$ compared to the LAPACK
implementation. The achieved flop rate is more than 52\% of the peak performance
of the KNL, while the LAPACK implementation achieves only 40\% of the peak.

For the performance analysis using  GPUs, a Haswell system with 8 NVIDIA K80s is
tested.  Figure~\ref{fig:time-falcon} shows the scalability with different numbers of
GPUs units. Using 1, 2, 4, 8, and 16 GPUs, we achieve an average
of 1.1, 1.9, 3.1, 5.2, and 6.6 Tflop/s, respectively. With this high flop rate, one
iteration of a 100 K problem can be solved using 16 GPUs in less than 52 seconds.

\begin{figure}
	\centering
	\includegraphics[width=0.9\linewidth]{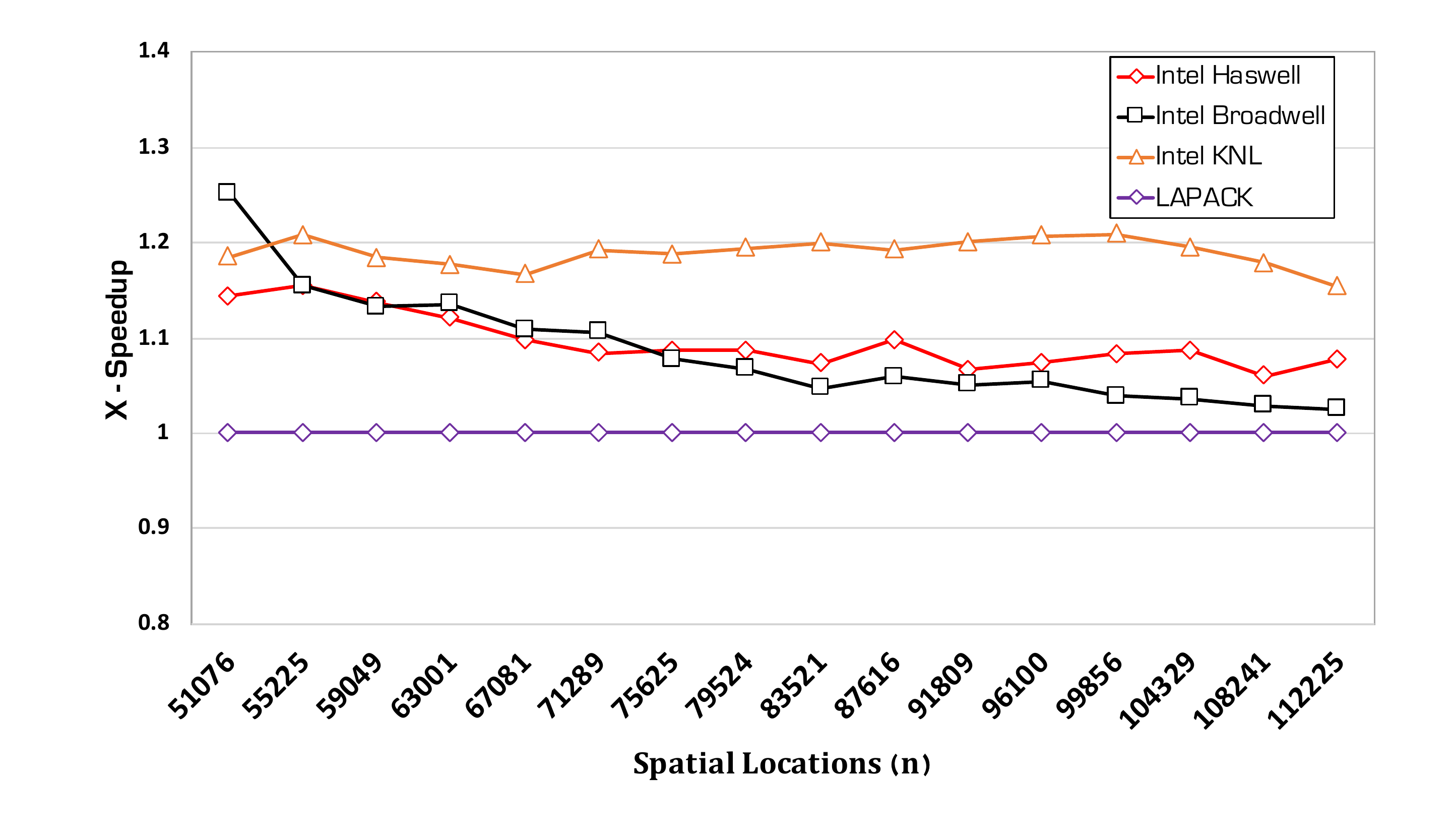}
	\caption{ExaGeoStat/Chameleon speedup compared to ExaGeoStat/LAPACK on Haswell, Broadwell, and KNL.}
	\label{fig:time-speedup}
\end{figure}

Figure~\ref{fig:time-speedup} shows the speedup gained from using 
 \emph{ExaGeoStat} based on Chameleon compared to LAPACK across a range of matrix sizes. The figure shows
the speedup based on the aforementioned Intel architectures, i.e., Haswell,  Broadwell, and KNL. The speedup can
reach up to 1.4$\times$, 1.25$\times$, and 1.2$\times$ on  these systems, respectively. The 
minimum gained speedup using the Haswell, Broadwell, and KNL processors  are  1.025$\times$, 1.06$\times$, and 1.15$\times$, respectively.
There is a performance trend. For small matrix
sizes, the asynchronous Chameleon-based \emph{ExaGeoStat} performs better than the traditional bulk 
synchronous LAPACK-based \emph{ExaGeoStat} 
since it mitigates the idle time between panel factorization 
and update of the trailing submatrix. For asymptotic matrix sizes, the performance gap shrinks between
both \emph{ExaGeoStat} variants since the workload is large enough to maintain 
hardware resources busy.

\begin{figure}
	\centering
	\includegraphics[width=0.9\linewidth]{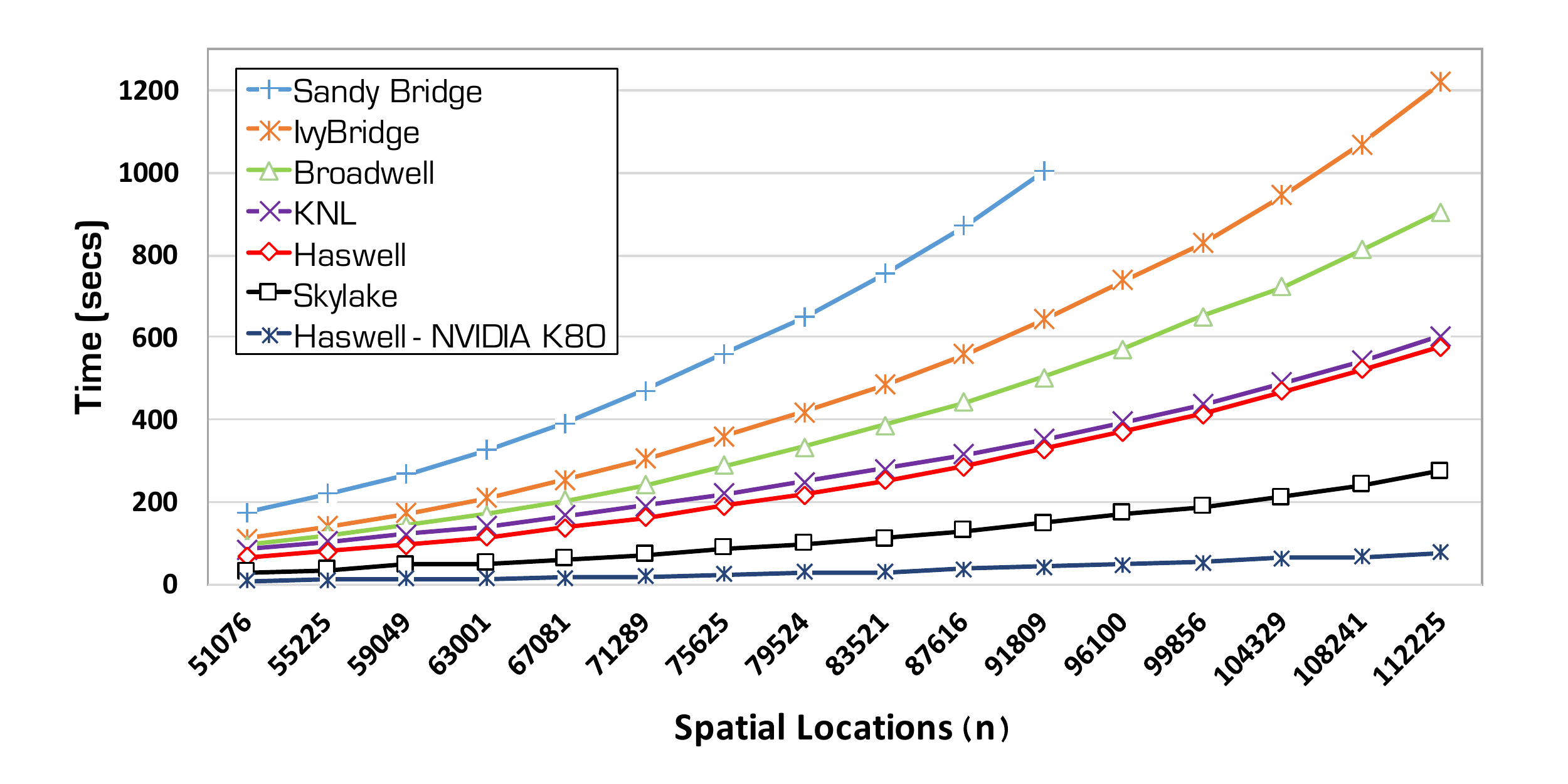}
	\caption{Hardware-agnostic \emph{ExaGeoStat} software.}
	\label{fig:time-intel}
\end{figure}

Figure~\ref{fig:time-intel} summarizes the performance of \emph{ExaGeoStat} on different shared-memory Intel processors. 
Skylake processor represents the latest available generation of Intel processors. Skylake
 shows the highest gained in terms of performance on x86 systems, 
 where $100$ K $\times$ $100$ K problems can be solved in about 4.5 minutes on $56$ cores. 
This experiment does not aim at comparing performance across Intel processor generations, 
since each system has a different total number of cores,
 i.e., Sandy Bridge ($32$ cores total), IvyBridge ($40$ cores total), 
 Broadwell ($28$ cores total), KNL ($64$ cores), 
 Haswell ($36$ cores total), Skylake ($56$ cores total), and eight NVIDIA K80 GPU server. 
 However, this experiment aims at showing how our proposed \emph{unified} software is 
 hardware-agnostic and can deploy on Intel x86 architectures as well as NVIDIA GPU-based
 servers, using a single source code, which is further leveraged 
 to distributed-memory environment systems. 

\begin{figure}
	\centering
	\includegraphics[width=0.9\linewidth]{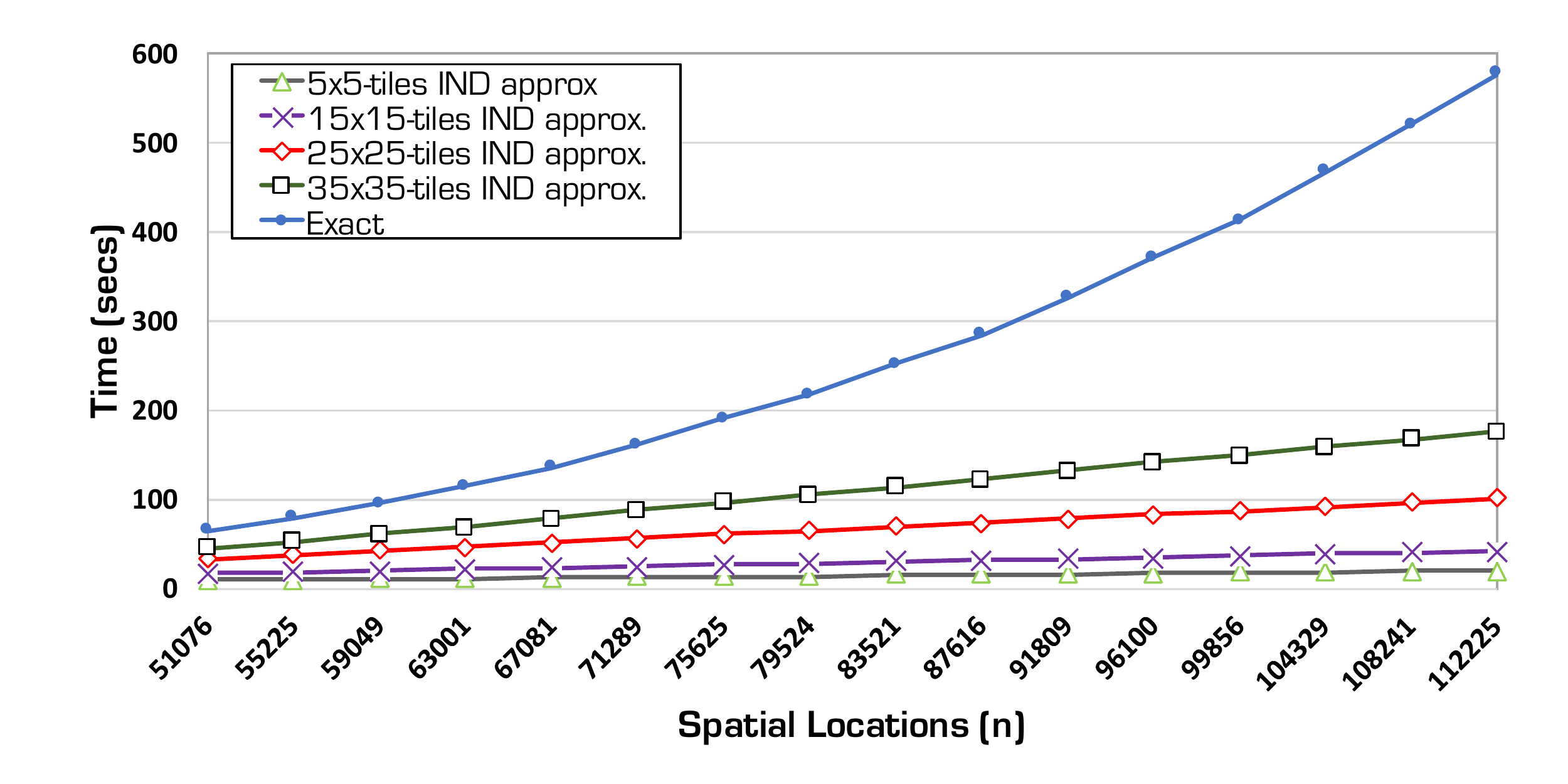}
	\caption{Time for one iteration of the likelihood estimation using exact and IND computation on Haswell processor.}
	\label{fig:shihab-time-approx}
\end{figure}

\begin{figure}[!ht]
	\centering
	\subfigure[One iteration of likelihood operation - time.]{
		\label{fig:time-shaheen}
	\includegraphics[width=0.85\linewidth]{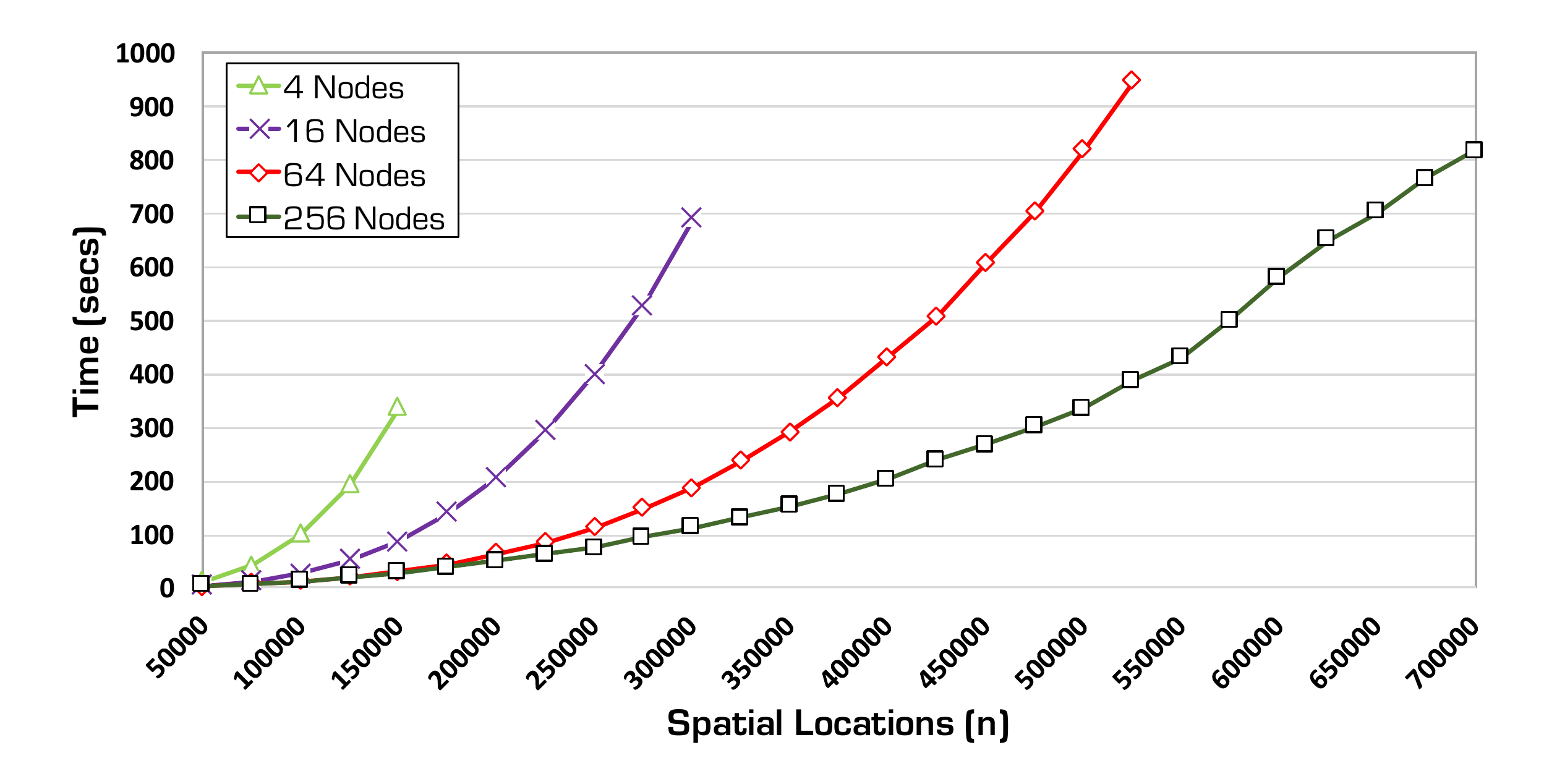}}
	\subfigure[One iteration of likelihood operation - Tflop/s.]{
		\label{fig:perf-shaheen}
	\includegraphics[width=0.90\linewidth]{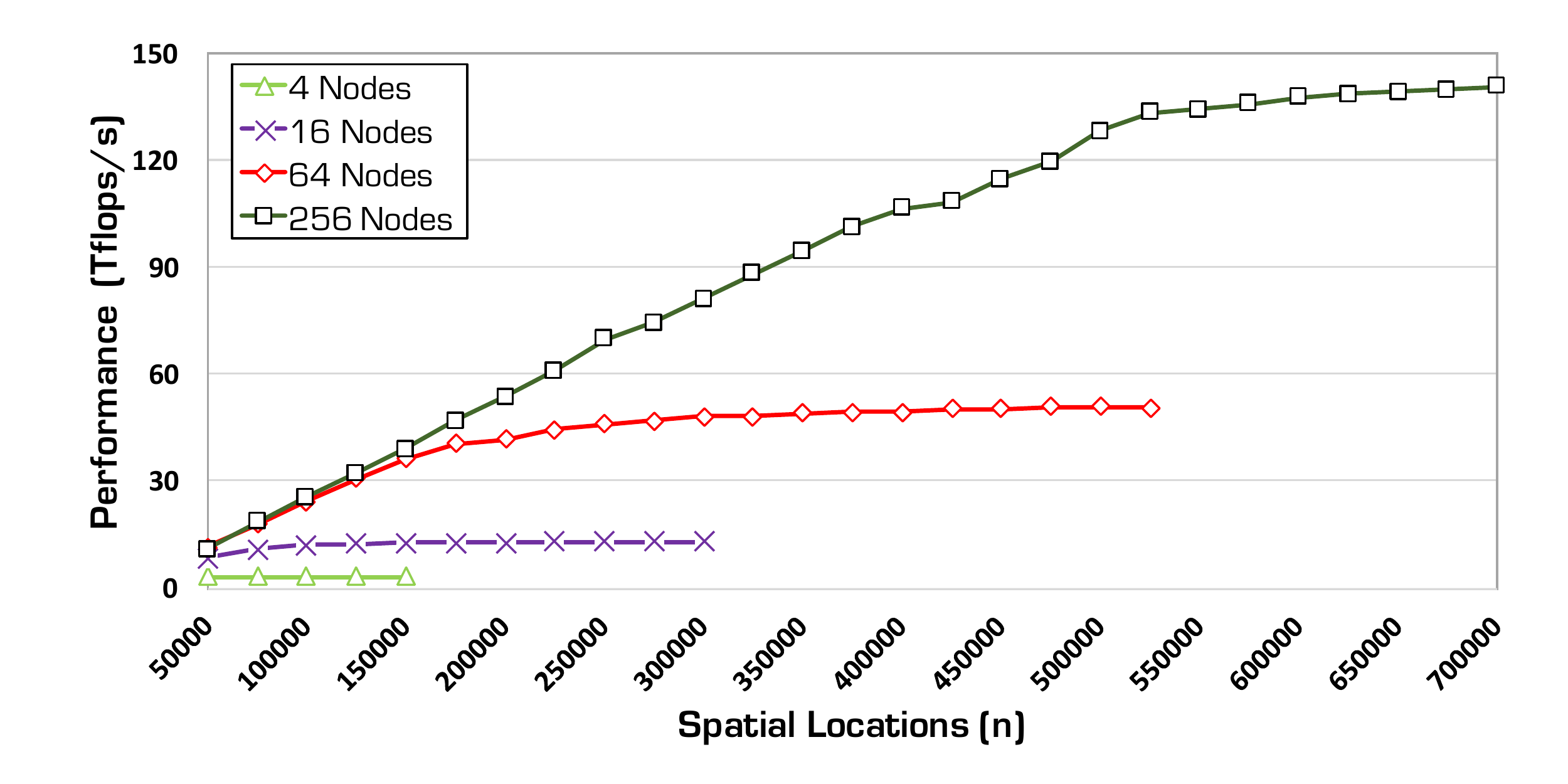}}
	\subfigure[Prediction eval. of 100 unknown values - time.]
	{
		\label{fig:time-shaheen2}
	\includegraphics[width=0.90\linewidth]{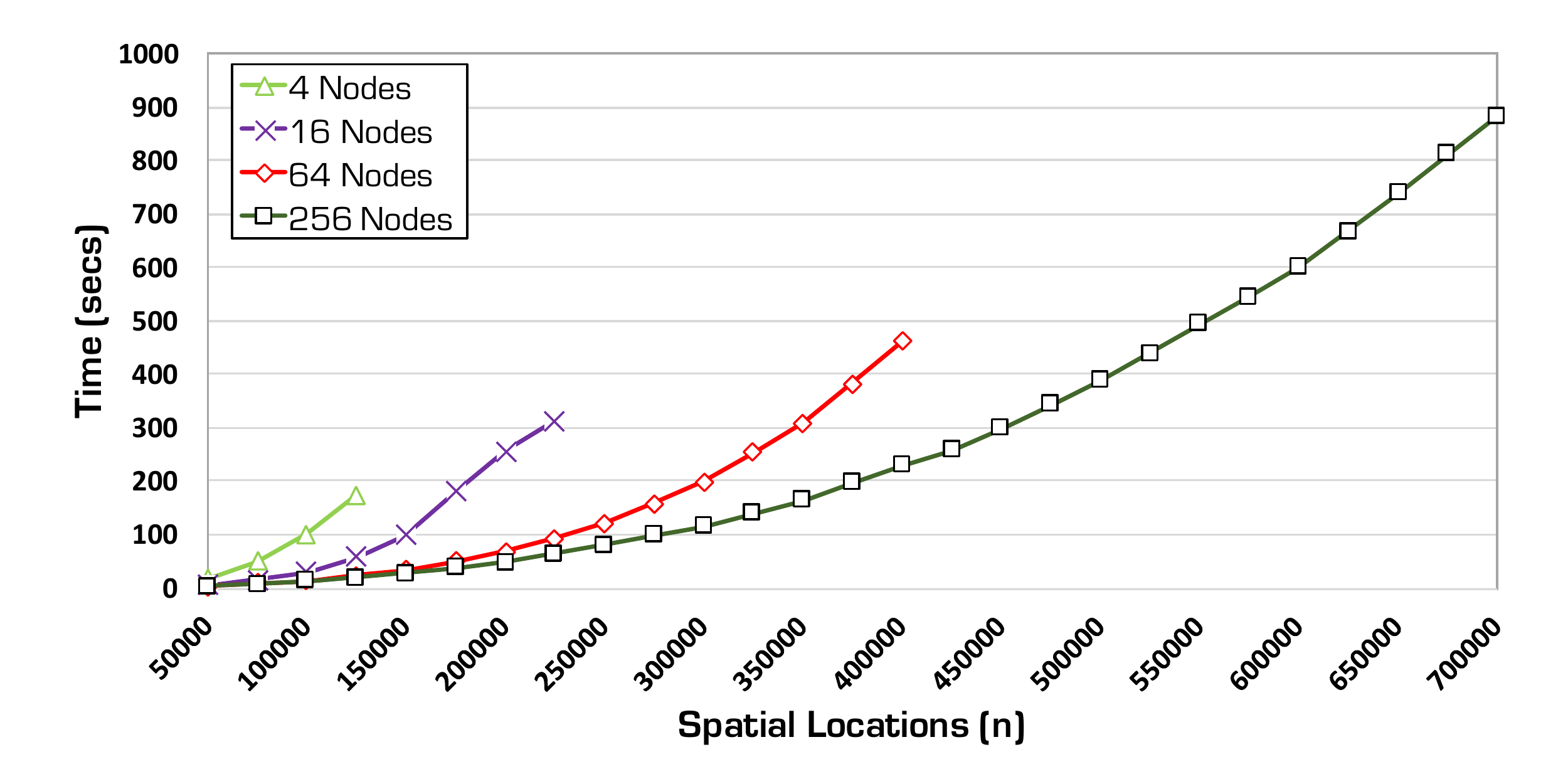}}
	\subfigure[Prediction eval. of 100 unknown values - Tflop/s.]
	{
		\label{fig:perf-shaheen2}
	\includegraphics[width=0.90\linewidth]{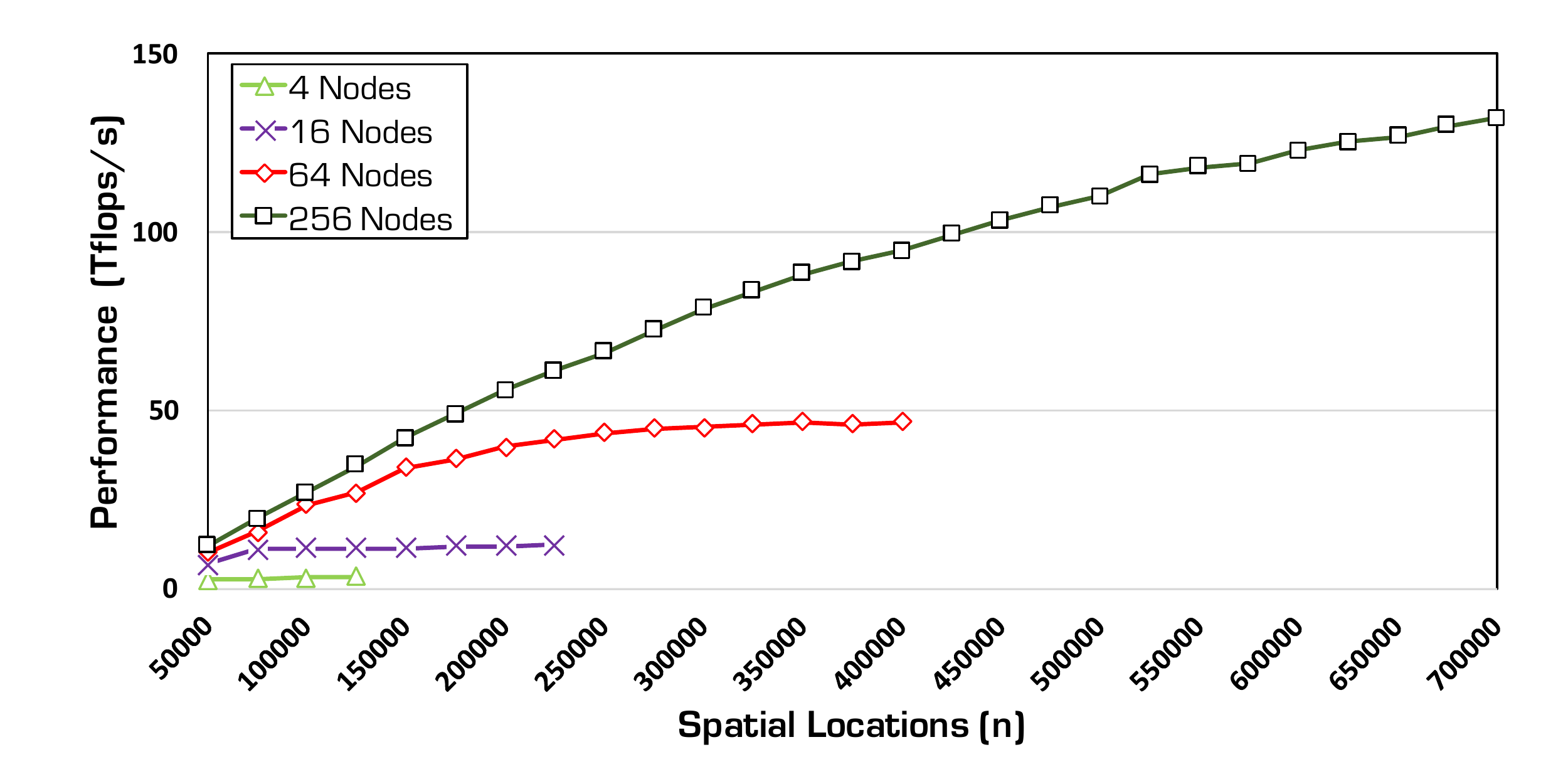}}

	\caption{Distributed-memory performance scalability on Cray XC40.}
	\label{fig:shaheen}
\end{figure}

\begin{figure*}[!ht]
	\centering
	\subfigure[Estimated variance parameter ($\theta_1$).]{
		\label{fig:theta1}
	\includegraphics[width=0.3\linewidth, height=0.16\textheight]{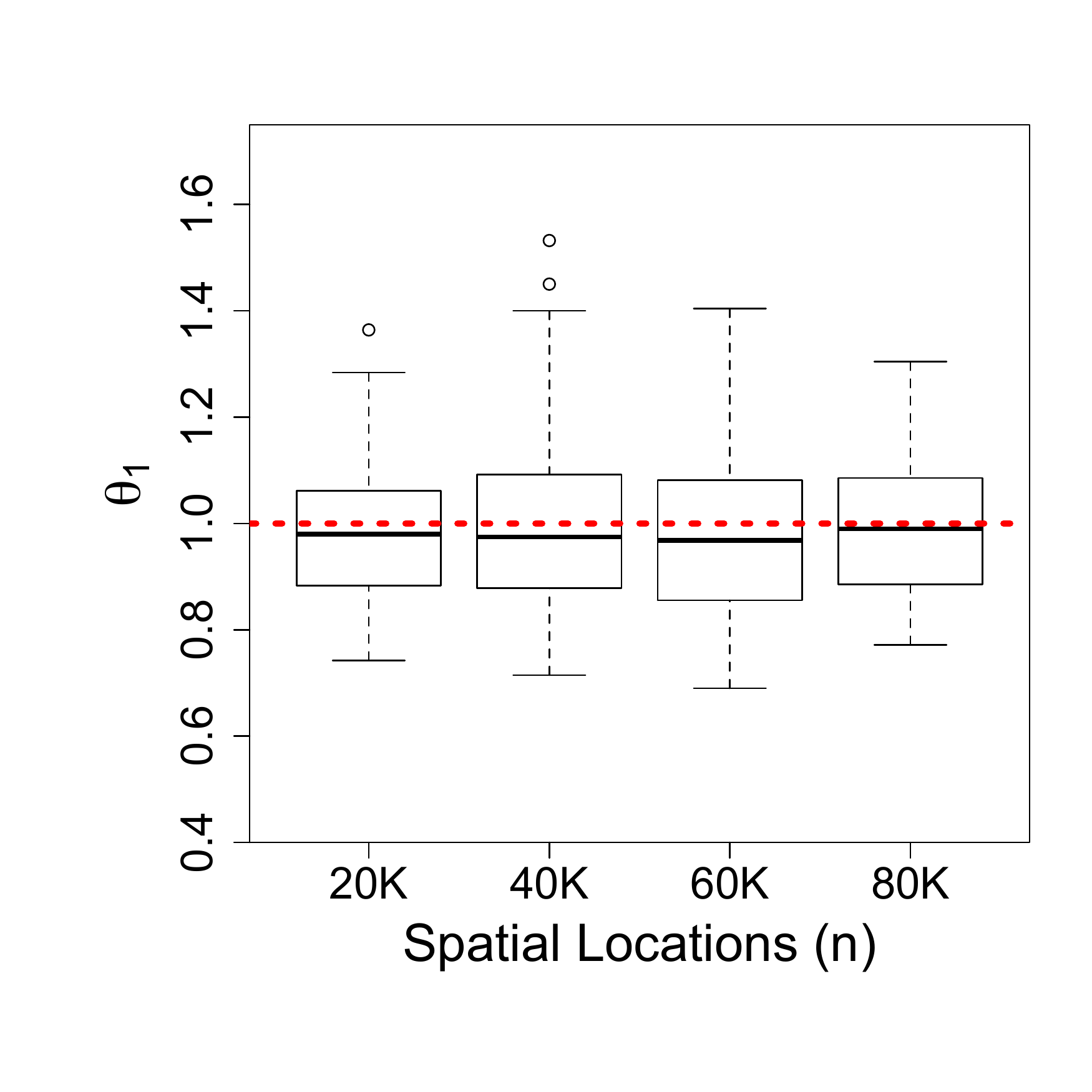}}
	\subfigure[Estimated spatial range parameter ($\theta_2$).]{
		\label{fig:theta2}
	\includegraphics[width=0.3\linewidth, height=0.16\textheight]{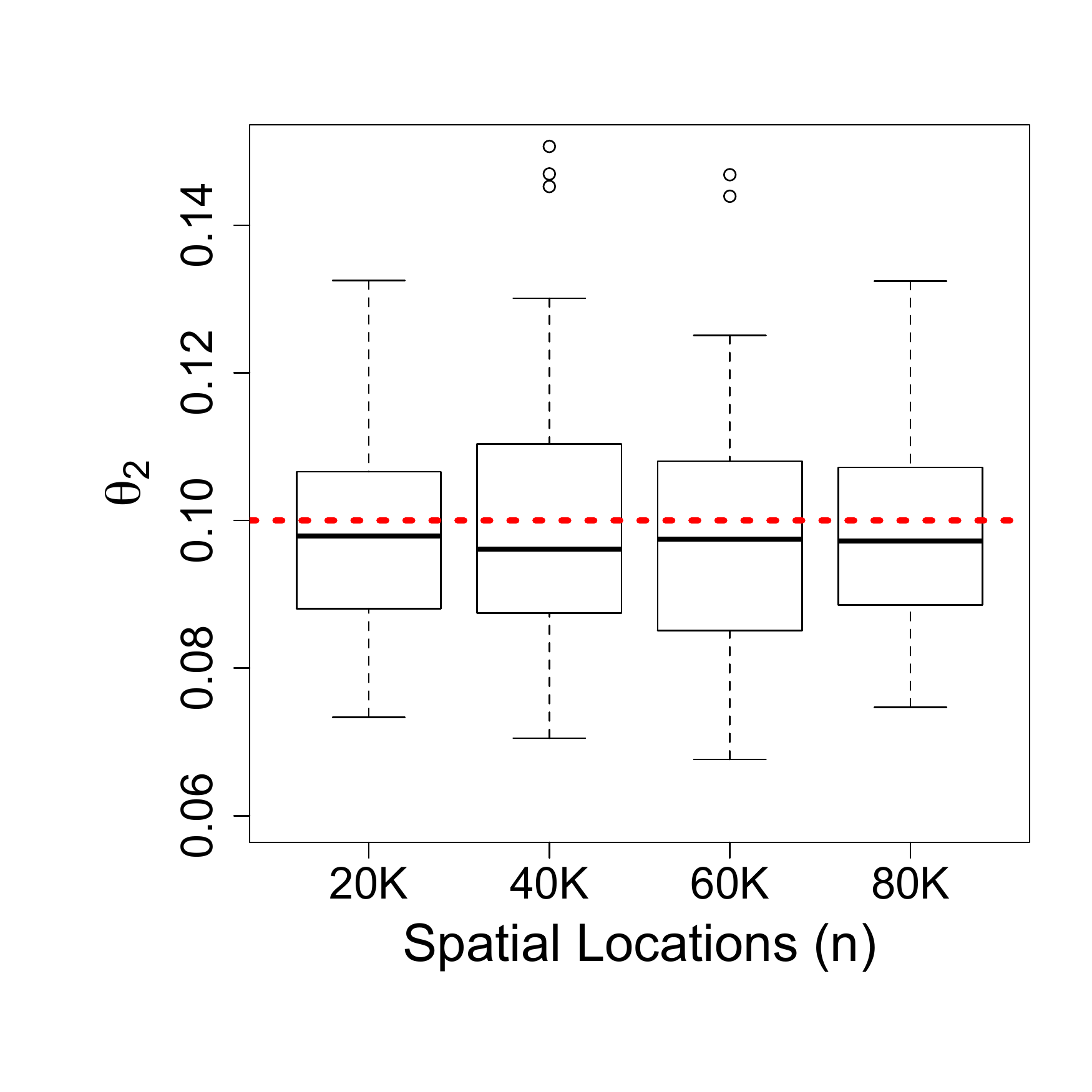}}
	\subfigure[Estimated smoothness parameter ($\theta_3$).]{
		\label{fig:theta3}
	\includegraphics[width=0.3\linewidth, height=0.16\textheight]{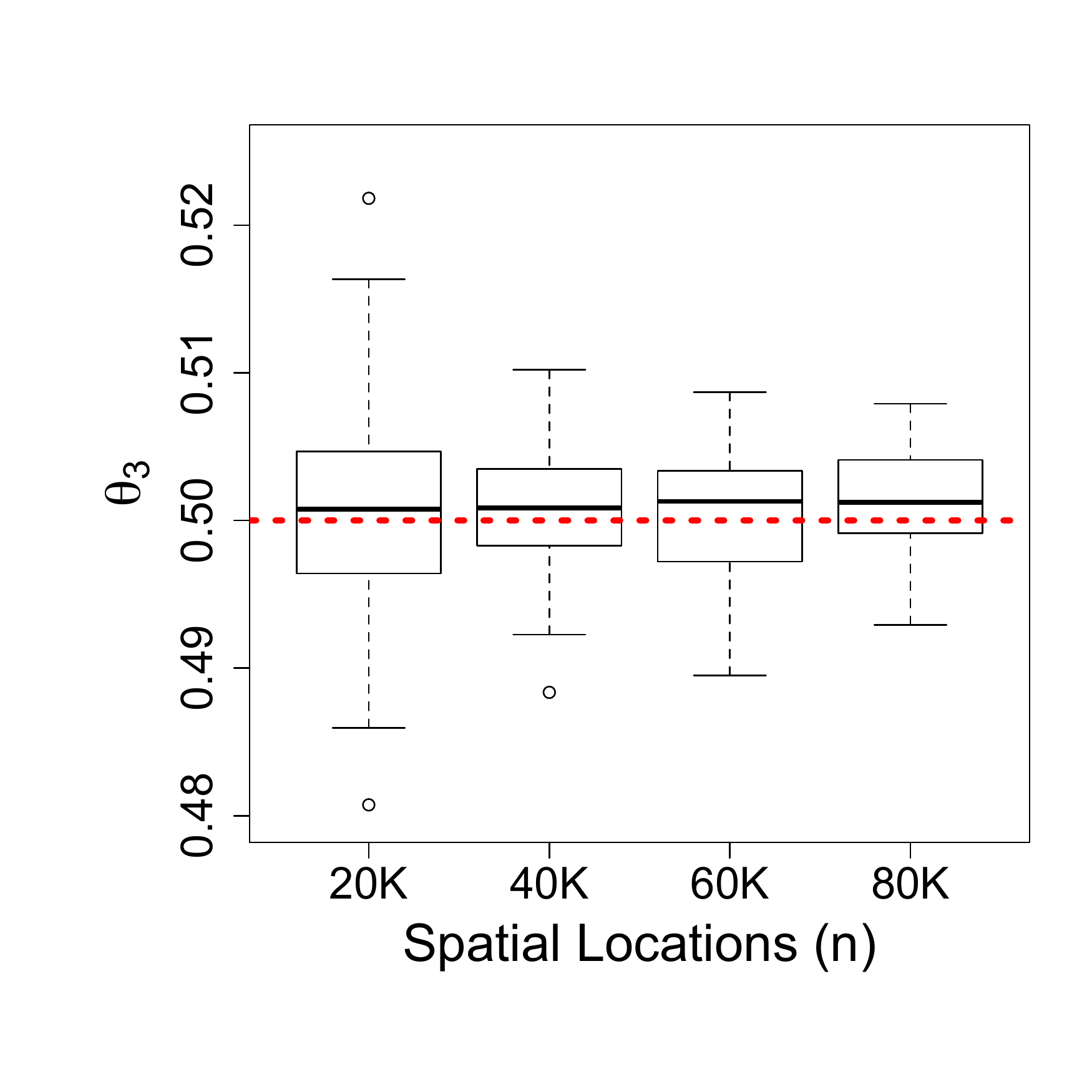}}
	\caption{Boxplots of parameter estimates. 
	}
	\label{fig:estimates}
\end{figure*}

Figure~\ref{fig:shihab-time-approx} shows the performance impact when
applying the IND approximation technique on synthetic datasets. 
We only report the performance on Intel Haswell processor, since similar performance trend
may be obtained on the other hardware systems. As expected, the IND approximation method 
is able to estimate the likelihood function faster than the exact method. The figure also shows 
that with larger diagonal super tiles, the likelihood estimation time increases, as expected. 
These performance numbers have to be cautiously put in the context of the qualitative study, 
presented later in Section~\ref{subsec:ind_approx}.

We also test our proposed software on the distributed-memory Cray XC40, 
\emph{Shaheen}, with different numbers of nodes. Figure~\ref{fig:time-shaheen}
reports the total execution time in terms of cores (with $32$ cores per node). 
With small matrix sizes, the benefits are modest.
However, as the matrix size grows, speedup saturates at higher and higher values.
\emph{ExaGeoStat} is able to solve one maximum
likelihood problem of dimension $700$ K in about $800$ seconds.

Figure~\ref{fig:perf-shaheen} shows the performance using the distributed-memory 
by reporting the flop rate against the varying core count. 
Using $8192$ cores, $140$ Tflop/s is achieved on a problem of dimension $700$ K. 
These experiments not only validate
the good performance of our unified platform on different hardware 
architectures, but also, to the best of our knowledge, extend the exact solution of the MLE problem
to such unprecedented large sizes. In Figures~\ref{fig:time-shaheen} and~\ref{fig:perf-shaheen}, 
some lines do not extend very far because of memory limits for smaller numbers of cores.
\subsubsection{Prediction Evaluation Performance}
Here, we investigate the performance of the prediction operation (i.e., 100 unknown measurements) using the \emph{ExaGeoStat} software on a distributed system (i.e., Cray XC40).

Figure~\ref{fig:time-shaheen2} shows the execution time for the prediction from different sizes synthetic datasets up to 700 K using 4, 16, 64, and 256 Shaheen's nodes, each has 32 cores. The scalability can seen in the figure. The prediction operation for a 700 K problem size can be evaluated in about 880 seconds.

Figure~\ref{fig:perf-shaheen2} shows the performance in Tflop/s with different numbers of cores. On Shaheen, the prediction operation achieve a 130 Tflop/s performance for 700 K problem size using 8192 cores.





\subsection{Qualitative Analysis (Monte Carlo Simulations)}

The overall goal of the maximum likelihood model is to estimate the unknown parameters of the statistical model  ($\theta_1$, $\theta_2$, and $\theta_3$) of the Mat\'{e}rn covariance function, then to use this model for future predictions of unknown measurements. In this experiment, we use the Monte Carlo simulation to estimate the parameters of an exponential covariance model, where  $\theta_1=1$, $\theta_2=0.1$, $\theta_3=0.5$.

Using our data generator tool and the initial parameter vector ($\theta_1=1$, $\theta_2=0.1$, $\theta_3=0.5$), four different synthetic datasets are generated (i.e., 20 K, 40 K, 60 K, and 80 K) besides 100 measurement vectors $\bf{Z}$ for each dataset. This experiment is repeated 100 times with different measurement vectors.

Figures \ref{fig:theta1},~\ref{fig:theta2}, and~\ref{fig:theta3} show three boxplots representing the estimated parameters with 100  measurement vectors $\mathbf{Z}$. The true value is denoted by a dotted red line. As shown, all of the results are close to  the correct ${\boldsymbol \theta}$ vector. 

\begin{figure}
	\centering
	\includegraphics[width=0.70\linewidth, height=0.2\textheight]{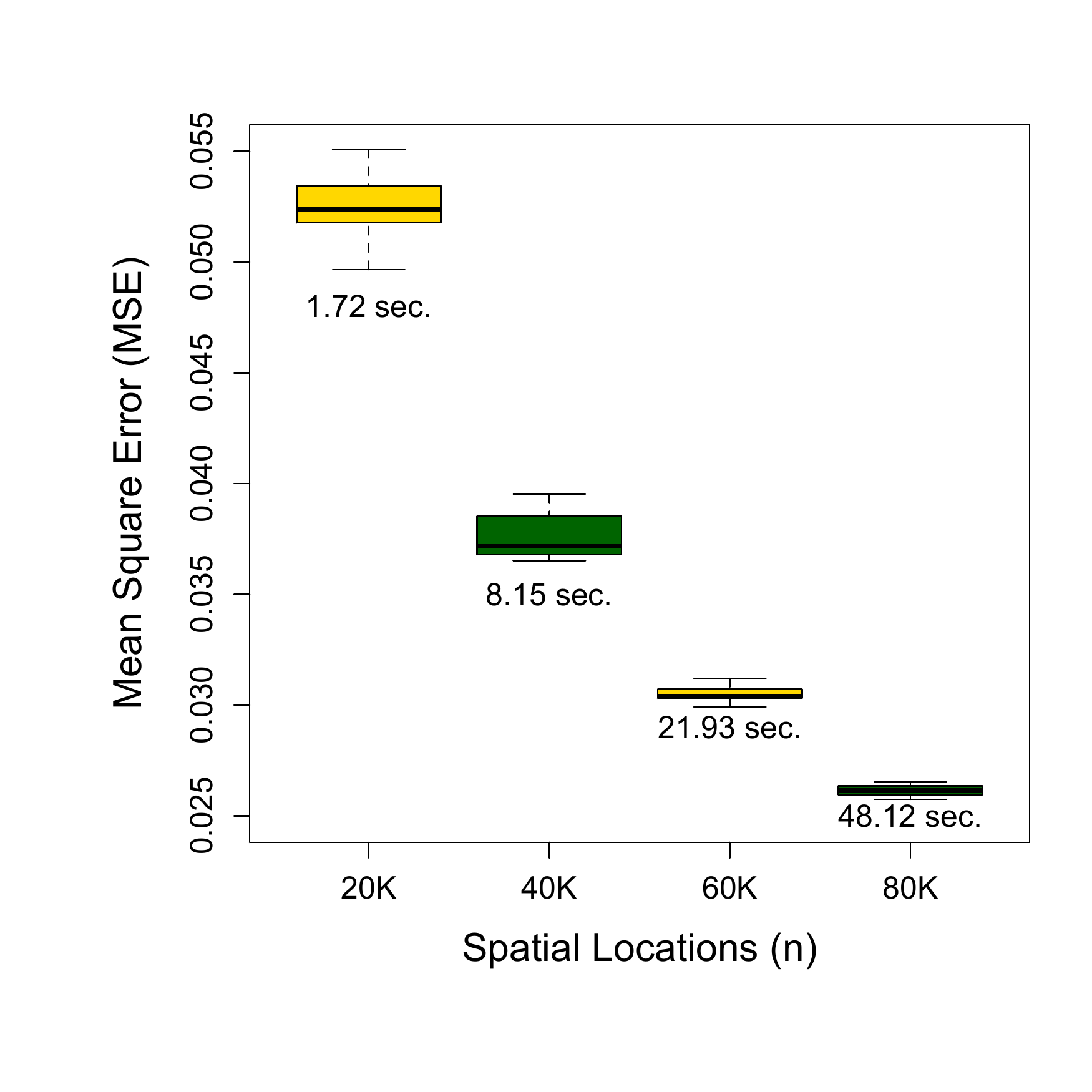}
	\caption{Predictions MSE of $n/k$ missing values of different synthetic dataset sizes using a $k$-fold cross-validation technique ($k$=10). Execution time per single prediction is shown under each boxplot using four nodes on Shaheen.}
	\label{fig:pred-estimates}
\end{figure}

To evaluate the accuracy of our predictions, we randomly choose a set of locations and mark the measurements on those locations as unknown. Then, using  the estimated ${\boldsymbol {\reallywidehat \theta}}$ vector, \emph{ExaGeoStat}  predicts the unknown measurements at those locations with the aid of the known measurements.  The accuracy of the prediction operation can be estimated using the Mean Square Error (MSE) between the actual measurements and the predicted ones as $MSE = \frac{1}{n} \sum_{i=1}^{n} (\reallywidehat{y_i} - y_i) ^2$,  where $\reallywidehat{y_i}$ represents the predicted value and $y_i$ represents the actual value at the same location.

Figure~\ref{fig:pred-estimates}  shows the boxplot of the predictions MSE using a $k$-fold cross-validation
technique, where $k$=10, to validate the prediction accuracy using
different synthetic dataset sizes. The total number of missing values equals to $n/k$ (i.e., subsample size). With larger matrix sizes, our prediction implementation has a smaller MSE compared to the smaller matrix sizes. The average execution time per single prediction using four nodes on Shaheen Cray XC40 is shown under each boxplot.

%
%
%


\begin{figure}
	\centering
	\includegraphics[width=9cm, height=4.6cm]{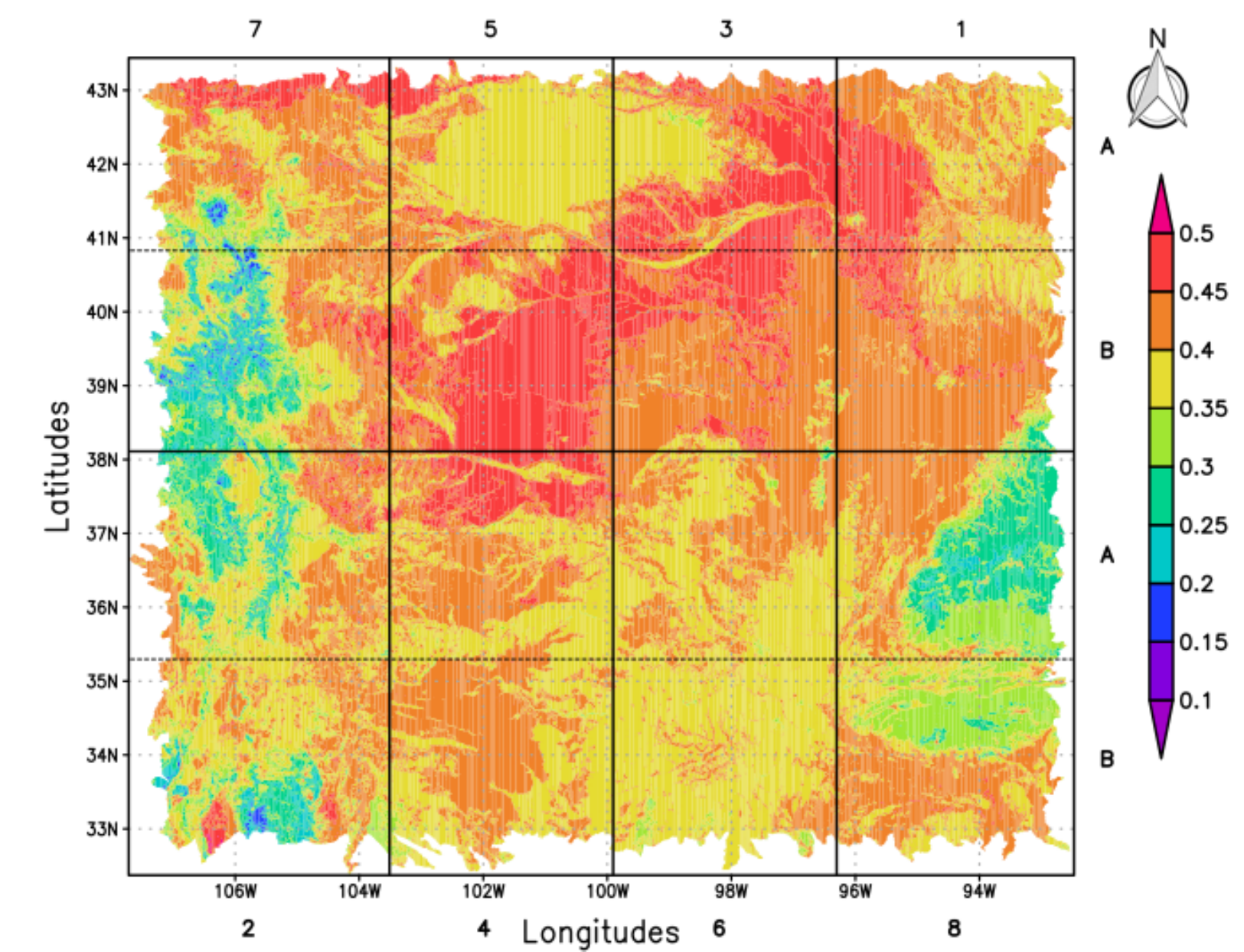}
	\caption{Soil moisture data divided into 16 geographical regions.}
	\label{fig:regions8}
\end{figure}

\subsection{Real Dataset Application}
In environmental applications, the number of measurements is usually very large.
These measurements are often distributed irregularly
across a given geographical region, and can be modeled as a realization from a Gaussian
spatial random field. In this study, we have evaluated our unified software using a soil
moisture data coming from Mississippi River basin region, USA (more details are given in Section 4). 
Because locations
in the soil moisture dataset are given by longitude and latitude pairs, the location space is 
non-Euclidean. Therefore, we use the Great-Circle Distance (GCD) metric 
to compute the distance between any given two locations with their original longitude and latitude values. The best representation of the GCD
is the haversine formula~\cite{rick1999deriving}
${\displaystyle \operatorname {hav} \left({\frac {d}{r}}\right)=\operatorname {hav} (\varphi _{2}-\varphi _{1})+\cos(\varphi _{1})\cos(\varphi _{2})\operatorname {hav} (\lambda _{2}-\lambda _{1})}$,\\
where hav is the haversine function, 
$\operatorname {hav} (\mathbf \theta )=\sin ^{2}\left({\frac {\mathbf \theta }{2}}\right)={\frac {1-\cos(\mathbf \theta )}{2}}$;
$d$ is the distance between two locations, $r$ is the radius of the sphere, 
$\varphi_1$ and $\varphi_2$ are the latitudes in radians of locations 1 and 
 2, respectively, and $\lambda _{1}$ and $\lambda _{2}$ 
are longitudes. 

For the soil moisture measurements from such a large spatial region, it is very likely 
that the process exhibits non-stationarity, i.e., 
the soil moisture covariance parameters may vary in space. Therefore, it is necessary to understand the features of the dataset before choosing
the appropriate statistical model for fitting the data. To examine whether it is reasonable to fit a 
stationary model to the whole spatial region, we propose dividing the 
entire region into disjointed subregions and applying our computationally efficient 
methods to fit stationary Gaussian process models with a Mat\'ern covariance 
function to each subregion. Then we can compare the spatial variability 
across regions using the parameter estimations. The division is only applied in order to study the behavior of the soil moisture dataset,
and does not mean that our method is limited
to stationary models. Our method can  be used directly on non-stationary covariance
models without modification. However, the stationary covariance models are essential
in any geospatial analysis and serve as the cornerstones of more complex, non-stationary models.

We consider two different strategies for dividing this dataset, 
as shown in Figure~\ref{fig:regions8}, where the locations are divided 
into 16 subregions (i.e., 1A, 1B,...etc.) or 8 subregions (i.e., 1, 2, ...etc.) The parameter estimation 
of the Mat\'ern covariance is summarized in Tables 1 and 2 using the Great-Circle Distance  (GCD). 

\begin{table}
\centering
	\setstretch{0.70}
\footnotesize
\label{tab:regions8}
\caption{Estimation of the Mat\'{e}rn covariance parameters for 8 geographical regions and the average prediction MSE using a $k$-fold cross-validation technique ($k$=10).}
\renewcommand{\arraystretch}{1.4}
\begin{tabular}{|c|c|c|c|c|}
\hline
& \multicolumn{3}{|c|}{Mat\'{e}rn Covariance}&  Avg. \\
Regions & Variance  &  Spatial   &  Smoothness &  prediction\\  
&   ($\theta_1$) &   Range ($\theta_2$) &  ($\theta_3$)  &  MSE\\ 

\hline

R 1& 0.823& 7.215& 0.529 &0.0643 \\ \hline
R 2& 0.481& 10.434& 0.500 &0.0315 \\ \hline
R 3& 0.328& 10.434& 0.534 &0.0175  \\ \hline
R 4& 0.697& 16.761& 0.483 &0.0298 \\ \hline
R 5& 1.152& 13.431& 0.482 &0.0612 \\ \hline
R 6& 0.697& 16.095& 0.512 &0.0263 \\ \hline
R 7& 0.520& 16.872&0.487&0.0213 \\ \hline
R 8& 0.390& 12.321& 0.447&0.0287 \\ \hline

	\end{tabular}
\end{table}

\begin{table}
	\centering
		\setstretch{0.9}
	\footnotesize
	\label{tab:regions16}
	\caption{Estimation of the Mat\'{e}rn covariance parameters for 16 geographical regions and the average prediction MSE using a $k$-fold cross-validation technique ($k$=10).}
	\renewcommand{\arraystretch}{1.4}
	\begin{tabular}{|c|c|c|c|c|}
		\hline
		&   \multicolumn{3}{|c|}{Mat\'{e}rn Covariance}& Avg.  \\
		Regions &  Variance  &  Spatial   &  Smoothness &  prediction.\\  
																																					 &   ($\theta_1$) &   Range ($\theta_2$) &  ($\theta_3$)  &  MSE\\ 

		\hline

		R 1A & 0.852&5.994& 0.559 & 0.0711 \\\hline
		R 1B & 0.380& 10.434& 0.490 & 0.0527\\ \hline
		R 2A & 0.277&10.878&0.507 & 0.0202\\ \hline
		R 2B & 0.410&7.77& 0.527& 0.0303\\ \hline
		R 3A & 0.836& 9.213& 0.496 & 0.0619\\ \hline
		R 3B & 0.619& 10.323& 0.523 & 0.0355\\ \hline
		R 4A & 0.553& 19.203&0.508& 0.0186\\ \hline
		R 4B & 0.906&27.861& 0.461& 0.0298\\ \hline
		R 5A & 1.619& 17.205& 0.466 & 0.0775\\ \hline
		R 5B & 1.028& 24.531& 0.498 & 0.0296\\ \hline
		R 6A & 0.599& 25.197& 0.457 & 0.0219\\ \hline
		R 6B & 0.332& 12.432& 0.418 & 0.0294\\ \hline
		R 7A & 0.625& 7.659& 0.523 & 0.0463\\ \hline
		R 7B & 0.467& 9.324& 0.549 & 0.0244\\ \hline
		R 8A & 0.485& 12.654&0.464& 0.0313\\ \hline
		R 8B & 0.383& 13.875&0.477 & 0.0211\\ \hline

	\end{tabular}
\end{table} 


From both tables, we see that the marginal variance $\theta_1$ and the spatial range parameter $\theta_2$ change across 
regions, suggesting that the local variability shows obvious non-stationarity. 
However, the smoothness parameter 
$\theta_3$ hardly changes at all across different regions. In 
the future studies, we may merge the subregions with similar parameter estimates 
and fit one stationary model to that combined region, while investigating the covariances of those subregions with very different parameter estimates more carefully.

We also estimate the accuracy by validating the
 estimated model parameters using a prediction 
 evaluation process.  A $k$-fold cross-validation
 technique has been used, where $k=10$. In this case,
the number of missing values that have been 
chosen from large regions, i.e., 250 K, is 25000, and from
small regions, i.e., 125 K, is 12500. We applied the $k$-fold cross-validation technique
and reported the average MSE. 

  
  Although GCD may be one of the best representations of the distance between two points on the surface of a sphere such
  as the earth, we have also tried the Euclidean Distance (ED) metric for the soil moisture data, 
  after transforming soil moisture dataset to the Euclidean space. We have found that none of
   GCD and ED are uniformly better than the other and, therefore, have decided to only report
    GCD metric.


\subsection{Qualitative Analysis Using Real Application}

\label{subsec:ind_approx}
Evaluating the accuracy of exact computation compared to approximation techniques is necessary
to highlight the advantage of using such a computational-intensive software
to solve the likelihood estimation problem. As mentioned earlier, one of the main goals of 
this study is to build a benchmark software to validate existing or future approximation
techniques with large-scale data.
 
In this experiment, we compare exact computation with IND approximation, i.e., both are
provided by the \emph{ExaGeoStat} software. The experiment is performed on one
region of soil moisture dataset (1A). The selected region has 125 K measurements
with a location matrix of size 125 K $\times$ 125 K elements. To gain the best performance of \emph{ExaGeoStat}, 
we tune the tile size to be 560, with Cray XC40 cluster to optimize the performance.
In this case, the whole location matrix is divided to 224 tiles in each dimension
, i.e.,  total number of tiles = $224$ $\times$ $224$ = $50176$ tiles. In the case of IND approximation,
 a different size of diagonal super tiles is used, i.e., $20 \times 20$, $40 \times 40$, $60 \times 60$, and
  $80 \times 80$. Of course, large diagonal super tiles show more accuracy improvements because more locations
  are taken into account. The experiment has been repeated for 100 times, each aims at predicting 
different 100 missing values randomly selected from the same region.

\begin{figure}
	\centering
	\includegraphics[width=0.75\linewidth, height=0.2\textheight]{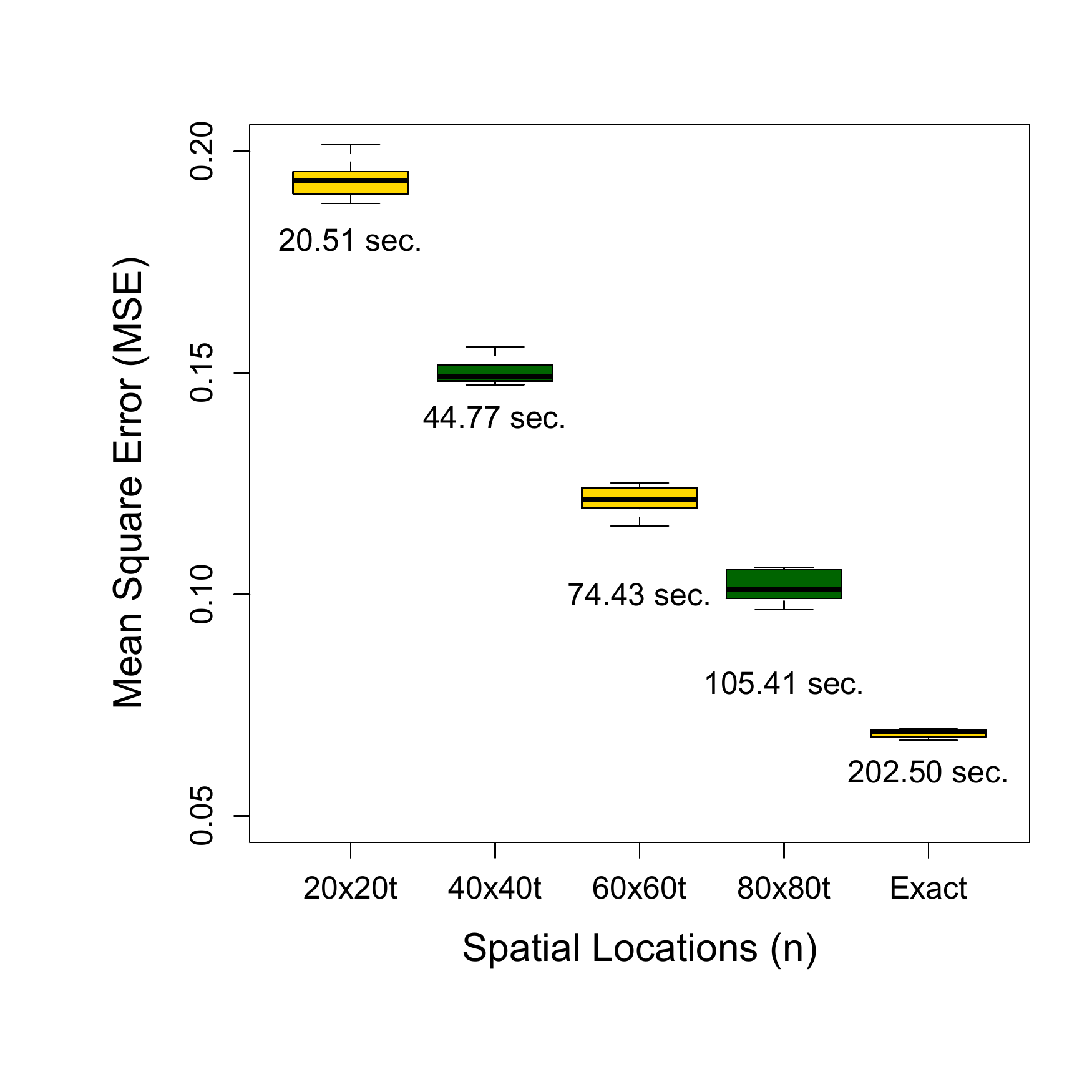}
	\caption{ Predictions MSE of 12500 missing values of region 1A of the soil moisture dataset using both exact and IND approx. methods with a $k$-fold cross-validation technique ($k$=10). Execution time per single prediction is shown under each boxplot using four nodes on Shaheen.}
	\label{fig:pred-estimates-ind}
\end{figure}

%
%
%
%
%
%
%

Figure~\ref{fig:pred-estimates-ind} shows the predictions MSE for various sizes 
of diagonal super tiles using IND approximation technique compared to the exact solution on region 1A.
 A $k$-fold cross-validation
technique, where $k$=10, is used to validate the prediction accuracy, where  the total number of missing values equals to 12500  (i.e., $n/k$ subsample).
As shown, exact computation satisfies the lowest MSE prediction compared to IND
approximation technique. The figure
also illustrates that lower MSE prediction values can be obtained by increasing the size
of diagonal super tiles in the case of IND approximation. In this case, 
more computation power is required to complete the whole estimation operation. Moreover, the
average execution time per single prediction on four Shaheen nodes is shown under each boxplot.

%

%
%
%
%


\section{Conclusion and Future Work}
\label{sec:summary}
This paper highlights the ability of the new \emph{ExaGeoStat} software to estimate the maximum likelihood function in the
context of climate and environmental applications and to predict missing measurements across geographical locations.
 This software provides a full machine learning pipeline (i.e., estimation, model fitting, 
and prediction) for geostatistics data.
\emph{ExaGeoStat}
is able to run with a decent performance on a wide range of hardware architectures, thanks to the high-performance, dense linear
algebra library Chameleon, associated with the StarPU runtime system.
We have successfully applied the software to synthetic and real
datasets.
Since calculations are performed without any approximations, the 
estimated parameters can be used as a references for assessing 
different approaches,with the end goal of generating online
database containing an ensemble of parameter estimates.

We also provide the implementation of an \emph{R} wrapper API, i.e., \emph{ExaGeoStatR}, 
to ease the process of integrating {\em ExaGeoStat} with the computational statistician community. 
The  \emph{R} package includes the main functions
that can be used by statisticians to evaluate the MLE operation on variant hardware architectures.

In the future work, we plan to investigate hierarchical matrix 
approximations based on $\mathcal{H}$-matrices, which will allow us to replace dense subblocks
of the exact matrix with low-rank approximations in an accuracy-tunable
manner, significantly reducing the memory footprint and operation count
without compromising the accuracy of the
applications~\cite{akbudak-isc17}. We also plan to support our package with NetCDF support to deal
with a wide range of existing climate and environment data.


\section{Acknowledgement}
\label{sec:ack}

The research reported in this publication was supported by funding from King Abdullah University of
Science and Technology (KAUST). We would like to thank NVIDIA for hardware donations in the context of a GPU Research Center, 
Intel for support in the form of an Intel Parallel Computing Center award, Cray for support 
provided during the Center of Excellence award to the Extreme Computing
Research Center at KAUST, and KAUST IT Research
Computing for their hardware support on the GPU-based system.
This research made use of the resources of the KAUST Supercomputing Laboratory.
Finally, the authors would like to thank Alexander Litvinenko from the Extreme Computing
Research Center for his valuable help.



\begin{IEEEbiography}
	[{\includegraphics[width=1in,height=1.25in,clip,keepaspectratio]{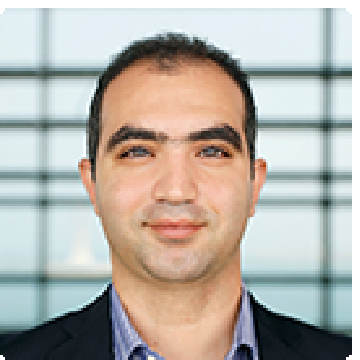}}]{Sameh Abdulah}
	is a Postdoctoral Fellow at the Extreme Computing Research Center, 
	King Abdullah University of Science and Technology, Saudi Arabia. Sameh
	received his MS and PhD degrees from Ohio State University, 
	Columbus, centered USA, in 2014 and 2016, His work is centered around High
	Performance Computing (HPC) applications, bitmap indexing in big data, large spatial datasets, parallel statistical applications, 
	algorithm-based fault tolerance,  and Machine Learning and Data Mining algorithms.
\end{IEEEbiography}


\vspace{-5mm}

\begin{IEEEbiography}
	[{\includegraphics[width=1in,height=1.25in,clip,keepaspectratio]{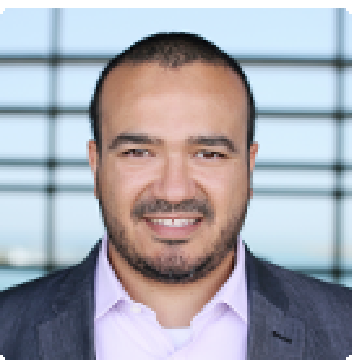}}]{Hatem Ltaief}
	is a Senior Research Scientist in the Extreme Computing
	Research Center at King Abdullah University of Science and Technology, Saudi Arabia.
	His research interests include parallel numerical
	algorithms, fault tolerant algorithms, parallel programming models,
	and performance optimizations for multicore architectures and hardware
	accelerators. His current research collaborators include Aramco,
	Total, Observatoire de Paris, NVIDIA, and Intel.
\end{IEEEbiography}

\begin{IEEEbiography}
	[{\includegraphics[width=1in,height=1.25in,clip,keepaspectratio]{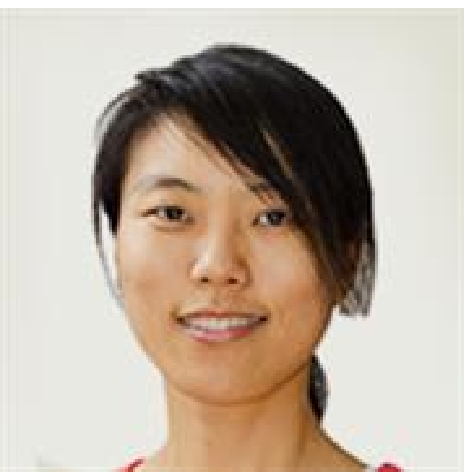}}]{Ying Sun}
	is an Assistant Professor of Statistics at King Abdullah University of Science and Technology (KAUST) in Saudi Arabia. She joined KAUST in June 2014 after one year of service as an assistant professor in the Department of Statistics at Ohio State University, USA. At KAUST, she leads a multidisciplinary research group on environmental statistics, dedicated to developing statistical models and methods for using space-time data to solve important environmental problems. Prof. Sun received her Ph.D. degree in Statistics from Texas A\&M University in 2011, and was a postdoctoral researcher in the research network of Statistics in the Atmospheric and Oceanic Sciences (STATMOS), affiliated with the University of Chicago and the Statistical and Applied Mathematical Sciences Institute (SAMSI). Her research interests include spatio-temporal statistics with environmental applications, computational methods for large datasets, uncertainty quantification and visualization, functional data analysis, robust statistics, and statistics of extremes.
\end{IEEEbiography}
\vspace{-5mm}
\begin{IEEEbiography}
	[{\includegraphics[width=1in,height=1.25in,clip,keepaspectratio]{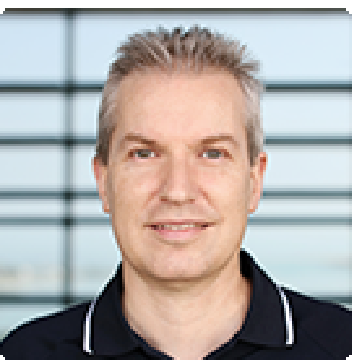}}]{Marc G. Genton}
	is a Distinguished Professor of Statistics at King Abdullah University of Science and Technology (KAUST) in Saudi Arabia. He received his Ph.D. in Statistics from the Swiss Federal Institute of Technology (EPFL), Lausanne, in 1996.  He also holds a M.Sc. in Applied Mathematics Teaching and a degree of Applied Mathematics Engineering from the same institution. Prof. Genton is a Fellow of the American Statistical Association, of the Institute of Mathematical Statistics, and the American Association for the Advancement of Science, and is an elected member of the International Statistical Institute. In 2010, he received the El-Shaarawi award for excellence from the International Environmetrics Society and the Distinguished Achievement award from the Section on Statistics and the Environment of the American Statistical Association. Prof. Genton's research interests include statistical analysis, flexible modeling, prediction, and uncertainty quantification of spatio-temporal data, with applications in environmental and climate science, renewable energies, geophysics, and marine science.
\end{IEEEbiography}
\vspace{-5mm}
\begin{IEEEbiography}
	[{\includegraphics[width=1in,height=1.25in,clip,keepaspectratio]{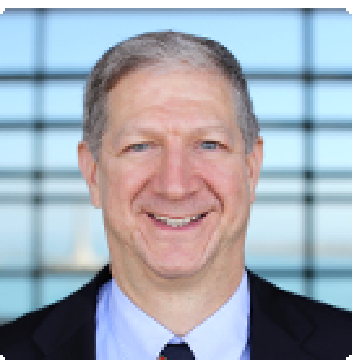}}]{David Keyes}
	directs the Extreme Computing Research Center at KAUST. He earned a BSE in Aerospace and Mechanical Sciences from Princeton University in 1978, and PhD in Applied Mathematics from Harvard University in 1984. Keyes works at the interface between parallel computing and the numerical analysis of PDEs, with a focus on scalable implicit solvers. He helped develop and popularize the Newton-Krylov-Schwarz (NKS), Additive Schwarz Preconditioned Inexact Newton (ASPIN), and Algebraic Fast Multipole (AFM) methods.  Before joining KAUST as a founding dean in 2009, he led multi-institutional projects researching scalable solver software in the SciDAC and ASCI programs of the US DOE, ran University collaboration programs at LLNL\textquotesingle s ISCR and NASA\textquotesingle s ICASE, and taught at Columbia, Old Dominion, and Yale Universities.  He is a Fellow of SIAM and AMS, and has been awarded the ACM Gordon Bell Prize, the IEEE Sidney Fernbach Award, and the SIAM Prize for Distinguished Service to the Profession.

\end{IEEEbiography}





\end{document}